# The Maxwell equations including magnetic monopoles

© W.D.Bauer        2004-1-28, 2004-2-28, 2005-4-10

email:             W.D.BAUER@t-online.de

**Abstract:**

The derivation of the Maxwell equations is reproduced whereby magnetic charges are included. This ansatz yields the results:

1) Longitudinal Ampère forces in a differential magnetostatic force law are improbable. Otherwise an electric current would generate magnetic charges.

2) Simple magnetic and electric induced polarization phenomena are completely analogous and are described by a Laplace equation.

3) Magnetic charges are the topological defects of a magnetic spin field similar like electric charges are the topological defects of an electric field. The magnetic charges should be interconnected with an elastic or inertial field which compensates the torque field generated by the magnetic moments of the anisotropic distributed spins.

4) Permanent magnetic fields can be understood to be caused by magnetic charges. Consequently, a moving permanent magnet represents a magnetic current which generates an electric field.

5) The electromagnetic tensors of energy and momentum have some additional terms which are written down generally.

6) Nonlinear electro-thermodynamic systems may violate the second law of thermodynamics. This is illustrated by an electric cycle with a data storing FET invented by Yusa & Sakaki.

# 1) Introduction

The Maxwell equations are about 150 Jahre old. They are the mathematical compilation of the experiments and considerations based on the original work of Cavendish, Coulomb, Poisson, Ampère, Faraday and others [1]. Mathematically they are partial differential equations. Different notations exist for them: most popular is the vector notation (O. Heaviside), which replaced the original notations in quaternions (J.C. Maxwell). More modern is the tensor notation (H. Minkowski, A. Einstein), which is able to describe situations which are discussed in the theory of relativity [2]. All notations are equivalent in the non-relativistic limit.

The Maxwell equations were and still are very successful. Until today their range of applicability grows permanently.

Here a short derivation is given which especially takes account for the newer developments of material descriptions. Furthermore, monopoles are included because Ehrenhaft proved their existence already 50 years ago [3-6]. It will be shown that the theory needs also their existence for a full description of all problems. This explains perhaps effects which are regarded generally as dubious because they cannot be understood in a conventional approach.

## 2) The equations of the electromagnetic field

a) The laws of Coulomb and the equation of Poisson

The so called Coulomb law describes the force between electric charges. It was discovered by Priestley in 1767 [1, 7]. Cavendish rediscovered it again and measured as well the dielectricity constant. However, due to many contributions to the knowledge about electricity it has the name of the third discoverer Couloumb [1].

The Coulomb law in the notation of today is [8]

$$F = \frac{1}{2}\sum_{i \neq j} q_i q_j \frac{(\mathbf{x_i}-\mathbf{x_j})}{|\mathbf{x_i}-\mathbf{x_j}|^3} = \iint \rho(\mathbf{x})\rho(\mathbf{x'}) \frac{(\mathbf{x}-\mathbf{x'})}{|\mathbf{x}-\mathbf{x'}|} \, d\mathbf{x'}^3 \, d\mathbf{x}^3 \qquad (1)$$

with the definitions F:=force, q:=single charge, x space coordinate and ´, i,j are indices. It can also be written as

$$F = \int \rho(\mathbf{x'})\mathbf{E}(\mathbf{x'}) \, d\mathbf{x'}^3 \qquad (2)$$

by using the definition of the electric field **E**

$$\mathbf{E} = \int \rho(\mathbf{x'}) \frac{(\mathbf{x}-\mathbf{x'})}{|\mathbf{x}-\mathbf{x'}|^3} \, d\mathbf{x'}^3 \qquad (3)$$

The electric field **E** can be derived from a potential $\Phi_E$ by using the definition

$$\mathbf{E} = -\nabla \Phi_E \qquad (4)$$

with the potential $\Phi_E$ defined by

$$\Phi_E = \int \frac{\rho(\mathbf{x}')}{|\mathbf{x}-\mathbf{x}'|} \, d\mathbf{x}'^3 \tag{5}$$

$\Phi_E$ has an empirical meaning. The **E**-field can be measured experimentally by a difference of voltage $\Phi_E(r) - \Phi_E(r_{ref})$ between a point in space at *r* and a reference point at $r_{ref}$ which oftenly is set to infinity where no field exists. Using the Poisson equation the charges of the field can be derived from the potential

$$\triangle \Phi_E = -\nabla \mathbf{E} = -4\pi\rho \tag{6}$$

If matter is in the field the empirical potential $\Phi_E$ consists of the induced charges $\rho_{matter}(\mathbf{x}')$ and the contributions from the charged surface of the conductors $\rho_{conductor}(\mathbf{x}')$

$$\Phi_E = \Phi_{conductor} + \Phi_P = \int (\rho_{conductor}(\mathbf{x}') + \rho_{matter}(\mathbf{x}')) \frac{1}{|\mathbf{x}-\mathbf{x}'|} \, d\mathbf{x}'^3 \tag{7}$$

where $\Phi_P$ is the "mean field" of the material charges. Per definition only the charges on the conductor are detectable experimentally. In order to obtain an expression with empirical variables similar to **(6)** the equation **(7)** is rewritten

$$\Phi_D := \Phi_E - \Phi_P = \Phi_{conductor} = \int \frac{\rho_{conductor}(\mathbf{x}')}{|\mathbf{x}-\mathbf{x}'|} \, d\mathbf{x}'^3 \tag{8}$$

Contrary to the empirical meaning of $\Phi_E$, $\Phi_D$ has only a formal character. Using $\Phi_D$ in the Poisson equation one can calculate the charges to be measured in or on the conductors. One defines

$$\mathbf{D} := \varepsilon_{ik}\mathbf{E} := \mathbf{E} + 4\pi\mathbf{P} := -\nabla\Phi_{\mathbf{D}} = \int \rho_{conductor}(\mathbf{x'})\frac{\mathbf{x}-\mathbf{x'}}{|\mathbf{x}-\mathbf{x'}|}\, d\mathbf{x'}^3 \quad (9)$$

where $\varepsilon_{ik}$ is the dielectric tensor of the material.

Using the mathematical relations $\nabla|\mathbf{x}-\mathbf{x'}|^{-1} = -\nabla'|\mathbf{x}-\mathbf{x'}|^{-1}$ and $\nabla'^2|\mathbf{x}-\mathbf{x'}|^{-1} = -4\pi\delta(\mathbf{x}-\mathbf{x'})$ and the redefinition $\rho := \rho_{conductor}$ the Poisson equation is

$$\Delta\Phi_{\mathbf{D}}(\mathbf{x}) = \nabla.[\varepsilon_{ik}(\mathbf{x})\nabla\Phi_{\mathbf{E}}(\mathbf{x})] = -4\pi\rho(\mathbf{x}) \quad (10)$$

Using **(9)** and **(10)** it follows generally

$$\nabla.\mathbf{D}(\mathbf{x}) = 4\rho(\mathbf{x}) \quad (11)$$

*Important special cases:*

  *surface charges*

An electric potential can exist due to a surface density $\sigma$

$$\Phi_{\mathbf{D}} := \Phi_{\mathbf{conductor}} = \int \frac{\sigma_{conductor}(\mathbf{x'})}{|\mathbf{x}-\mathbf{x'}|}\, d\mathbf{x'}^2 \quad (12)$$

Then, the electric field **D** is

$$\mathbf{D} := -\nabla\Phi_{\mathbf{D}} = \int \sigma(\mathbf{x'})\frac{\mathbf{x}-\mathbf{x'}}{|\mathbf{x}-\mathbf{x'}|^3}\, d\mathbf{x'}^2 \quad (13)$$

  *constraints for the material properties*

In the most cases it is possible to make simplifying constraints for the material properties. In order to explain this it is necessary to write down the potential $\Phi_\mathbf{P}(\mathbf{x})$ of the multipole expansion of the charges in the material [9]. A multipole expansion of the potential calculates the distribution of charge in space about a origin 0 as a serie of moments

$$\Phi_P(\mathbf{x}) = \frac{\rho}{r} + \frac{\mathbf{p}_{dipol} \cdot \mathbf{x}}{r^3} + \frac{1}{2} \sum_{i,j} Q_{ij} \frac{x_i x_j}{r^5} + \ldots \quad (14)$$

cf. appendix 1. Here the definitions of the dipol moment $\mathbf{p}_{dipol}$ and the quadrupol moment $Q_{ij}$ are

$$\mathbf{p}_{dipol} := \int \mathbf{x}' \rho(\mathbf{x}') dx'^3 \qquad Q_{ij} := \int (3 x'_i x'_j - r^2 \delta_{ij}) \rho(\mathbf{x}') dx'^3 \quad (15)$$

This consideration is done for all points in space. Using $\mathbf{p}_{dipol}$ as density of polarisation then follows

$$\Delta \Phi_P(\mathbf{x}',\mathbf{x}) = \left[ \frac{\rho(\mathbf{x}')}{|\mathbf{x}_i - \mathbf{x}'_i|} + \frac{\mathbf{P}_{dipol}(\mathbf{x}') \cdot (\mathbf{x}-\mathbf{x}')}{|\mathbf{x}_i - \mathbf{x}'_i|^3} + \frac{1}{2} \sum_{i,j} Q_{ij}(\mathbf{x}') \frac{(x_i - x'_i)(x_j - x'_j)}{|\mathbf{x}_i - \mathbf{x}'_i|^5} + \ldots \right] dV \quad (16)$$

The first term represents induced charges for instance if recombination processes in semiconductors have to be accounted for. For the most problems, however, electric neutrality can be assumed and the first term becomes zero. Furthermore oftenly higher terms are neglected because they are quantitatively irrelevant. Then, after integration over the whole space holds

$$\Phi_P(\mathbf{x}) = \int \mathbf{P}_{dipol}(\mathbf{x}') \cdot \frac{(\mathbf{x}-\mathbf{x}')}{|\mathbf{x}_i - \mathbf{x}'_i|^3} dx'^3 = -\int \frac{\nabla' \mathbf{P}_{dipol}(\mathbf{x}')}{|\mathbf{x}_i - \mathbf{x}'_i|} dx'^3 \quad (17)$$

If **(17)** is inserted in **(8)** one can identify: $\mathbf{P}=\mathbf{P}_{dipol}$, meaning, for this special case **P** can be identified with polarization.

b) Ampère´s law

The discovery of electromagnetism by Oersted [10, 11] in 1820 inspired some researchers in France to find the quantitative laws of these effects. Especially, Biot&Savart and Ampère tackled the task to solve this problem by intelligent experiments [1, 11]. In order to fit their experiment by a theory they made additional assumptions which filled up some lacking observations. This led to different laws for the forces between differential current elements of a circuit. For closed circuits, however, the different versions coincided in one law. The discussion of this problem is running until today.

Biot and Savart [12-14] found out that "the total force which is exerted by a file of infinite lenght under current on an element of austral or boreal magnetism in the distance FA or FB, is perpendicular on the shortest distance between the molecule and the file (see figs.1)". This law is written today in a form which goes back to Grassmann [11, 15]. It holds [8]

$$d\mathbf{F} \sim i_1 . i_2 \frac{d\mathbf{s_1} \times (d\mathbf{s_2} \times \mathbf{r}).}{|\mathbf{r}|^3} = i_1 . i_2 \left[ -(d\mathbf{s_1} . d\mathbf{s_2}) \frac{\mathbf{r}}{|\mathbf{r}|^3} + \left( \frac{d\mathbf{s_1} . \mathbf{r}}{|\mathbf{r}|^3} \right) d\mathbf{s_2} \right] \quad (18)$$

with $i_{1/2}$ := current, $d\mathbf{s}_{1/2}$ := length of file element, $\mathbf{r}$ := distance between file elements.

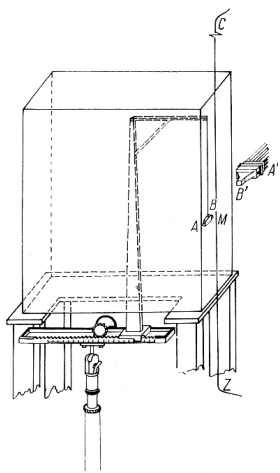 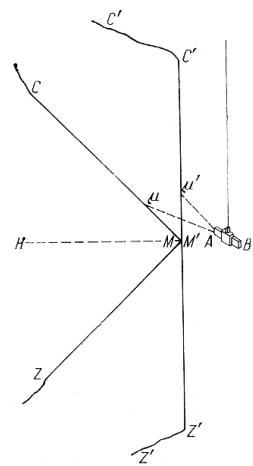

**fig.1a: the Biot-Savart - setup**
A magnetic needle is under the influence of the field of current CZ. A cover protects against the movement of air. The magnet A'B' compensates the magnetism of earth where the needle is located.

**Fig.1b: the Biot-Savart - setup**
measuring the time constant of the torsion pendulum it is concluded on the force of the field on the needle, if the current flows. Distance and angle of the file are varied in the experiments.

Ampère's law based mainly on the four following important observations [16, 17] which he extracted by his investigations:

1) The force of a file under current reverses if the current reverses, see fig.2a.

2) the forces of a current, which flows in a smooth circular circuit, is the same, if the "circle" of the current is not smooth but sinoidal, see fig.2b.

3) the force of the field of a closed current on a single current element is perpendicular to it, see fig.2c .

4) the force between two current elements does not change if all spatial dimensions of the setup are enlarged by a constant factor, see fig. 2d.

Additionally, in the tradition of Newton, he made the assumptions 1) that Newton's 3.axiom (actio-reactio) of mechanics also holds for electromagnetism, and 2) that the force between single elements of current is a central force which points into the direction of the shortest distance line between the elements.

Applying these ingredients Ampère constructed his force law. Based on Ampère´s assumption it holds for the force $\mathbf{F} \sim \mathbf{r}$. The observations 1)+ 2) suggest for first order

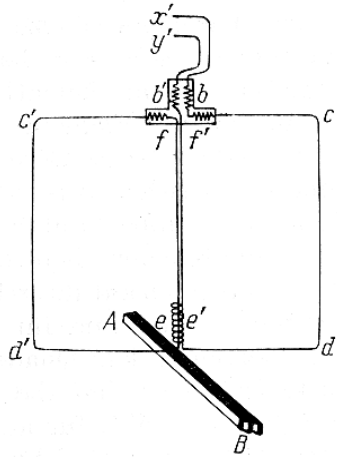
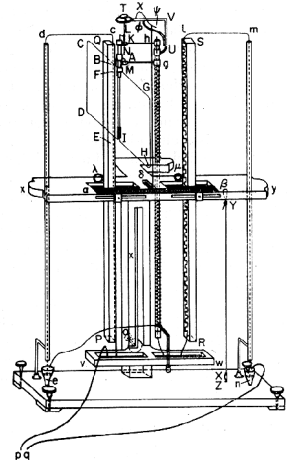

**fig.2a: Ampère´s first experiment**
AB is a fixed conductor under current. The circuits d'c'fe and cde'f' are stiffly connected. They are symmetrical over AB and can rotate about the axis x'y'. Their orientation of the current is opposite in these circuits; experimental result: no rotation due to complete balance of opposite forces if x'y' meets the middle point of AB

**fig.2b: Ampère´s second experiment**
In the trench PQ a current flows straight on in a conductor, in the trench SR it flows in a sinoidal conductor. The circuits BCDE and FGHI are mounted stiffly together, but can rotate around the Axis AK . The same current flows through them, however in opposite direction.
experimental result: only if the circuit is exaxtly in the middle between the conductors all forces compensate and no movement is observable.

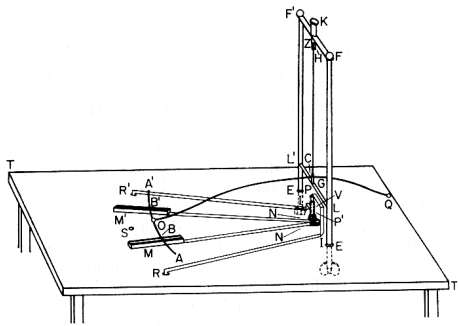
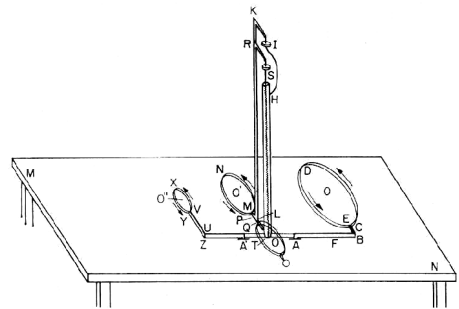

**fig.2c: Ampère´s third experiment**
M and M' are trenches filled with mercury, arm OC can be turned. The current flows over the troughs M back to the arm OC. The arm turns into the middle, where an equilibrium of torque exists and where all forces on OC apply perpendiculary.

**fig.2d: Ampère´s fourth experiment**
the outer circuits are fixed, the circuit in the middle can move. Only, if the diameters fulfil the relation $d_{left} : d_{middle} = d_{middle} : d_{right}$, all forces compensate and the circuit in the middle NOM does not move.

$\mathbf{F} \sim i_1.i_2[\varphi(r).(d\mathbf{s}_1.d\mathbf{s}_2) + \psi(r)(d\mathbf{s}_1.\mathbf{r}).(d\mathbf{s}_2.\mathbf{r})]$, the combination of both proportionalities result in $\mathbf{F} \sim i_1.i_2\mathbf{r}[\varphi(r).(d\mathbf{s}_1.d\mathbf{s}_2) + \psi(r)(d\mathbf{s}_1.\mathbf{r}).(d\mathbf{s}_2.\mathbf{r})]$. Observation 4) implies $\varphi(r)=A/r^3$ and $\psi(r)=B/r^5$ with A and B as constants to be determined. These can be calculated applying observation 3) as shown in the proof below. So follows B = -3A/2.

Proof[1]:

Imagine two circuits located with an angle of 90° between. Due to observation 3) and Ampère's assumption it holds for the force of a closed circuit of $\mathbf{s}_1$ on $d\mathbf{s}_2$

$$\mathbf{F}.d\mathbf{s}_2 \sim \sum_{\mathbf{s}_1} i_1.i_2 \mathbf{r}\left[\frac{A}{r^3}(d\mathbf{s}_1.d\mathbf{s}_2) + \frac{B}{r^5}(d\mathbf{s}_1.\mathbf{r}).(d\mathbf{s}_2.\mathbf{r})\right]d\mathbf{s}_2 = 0$$

Because this integral over the circuit $\mathbf{s}_1$ is zero, the integrated function is a potential with respect to $\mathbf{s}_1$. It can be expressed as well as a total differential

$$\frac{A\,(d\mathbf{s}_1.d\mathbf{s}_2).(d\mathbf{s}_2.\mathbf{r})}{r^3} + \frac{B(d\mathbf{s}_1.\mathbf{r}).(d\mathbf{s}_2.\mathbf{r})^2}{r^5}$$

Doing the integration one replaces $d\mathbf{s}_1$ by $d\mathbf{s}_1=-d\mathbf{r}$ and obtains

$$-\frac{A}{2r^3}\,d\left[(d\mathbf{s}_2.\mathbf{r})^2\right] + \frac{B(d\mathbf{s}_1.\mathbf{r}).(d\mathbf{s}_2.\mathbf{r})^2}{r^5}$$

Now, due to partial integration and due the potential property respect to $\mathbf{r}$ one can compare as coeffients

$$d\left(\frac{A}{2r^3}\right) = -\frac{B}{r^5}(\mathbf{ds}_1.\mathbf{r})$$

Using again $d\mathbf{s}_1=-d\mathbf{r}$ on the right side this becomes

$$-\frac{3A}{2r^4}dr = \frac{B}{r^4}dr$$

and B = -3A/2 follows. q.e.d □

So Ampère´s law is written :

$$\mathbf{F} = \frac{i_1 \cdot i_2}{c^2} \mathbf{r} [\frac{2}{r^3}.(d\mathbf{s_1}.d\mathbf{s_2}) - \frac{3}{r^5}(d\mathbf{s_1}.\mathbf{r}).(d\mathbf{s_2}.\mathbf{r})] \tag{19}$$

Riemann [18] and Whittaker [1] checked this derivation and realized, that Ampère's workout is only one possible ansatz to explain the observations. They doubted in Ampère's assumption, that the force between current elements is a central force, because the forces could be as well angular moments [19]. They found other possible formulas, which could explain all observations. Whittaker enlarged Ampère's formula and added terms, which were in accordance with the observations on closed current loops, because these additions were zero after integration over a closed loop. So he made the general ansatz:

$$\begin{aligned}\mathbf{F} = &-\frac{i.i'}{c^2}\mathbf{r}[\frac{2}{r^3}.(d\mathbf{s}.d\mathbf{s}') - \frac{3}{r^5}(d\mathbf{s}.\mathbf{r}).(d\mathbf{s}'.\mathbf{r})] \\ &+ \chi(r)(d\mathbf{s}'.\mathbf{r}).d\mathbf{s} + \chi(r)(d\mathbf{s}.\mathbf{r})d\mathbf{s}' + \chi(r).(d\mathbf{s}.d\mathbf{s}').\mathbf{r} \\ &+ \frac{1}{r}\chi'(r).(d\mathbf{s}.\mathbf{r}).(d\mathbf{s}'.\mathbf{r})\end{aligned} \tag{20}$$

Whittaker dropped Ampère's assumption, that the force should be a central force and he applied only Newton's law actio-reactio. He made the most simple possible choices for $\chi(r) = i.i'/(c^2 r^3)$ and $\chi'(r) = -3i.i'/(c^2 r^3)$ and obtained the force law

$$\mathbf{F} = \frac{i.i'}{c^2 r^3}[(d\mathbf{s}.\mathbf{r})d\mathbf{s}' + (d\mathbf{s}'.\mathbf{r})d\mathbf{s} - \mathbf{r}(d\mathbf{s}.d\mathbf{s}')] \tag{21}$$

Tabel 1: different versions of magnetostatic force law between current elements ([20] and [1])

general form of the magnetostatic force law:

$$F = k \frac{i.i´}{r^3}[\mathbf{r}.(A.(d\mathbf{s}.d\mathbf{s}´) + B.(\mathbf{r}.d\mathbf{s})(\mathbf{r}.d\mathbf{s}´)/r^2) + C.(\mathbf{r}.d\mathbf{s}´)d\mathbf{s} + D.(\mathbf{r}.d\mathbf{s})d\mathbf{s}´]$$

| name | year | ref. | A | B | C | D | comment |
|---|---|---|---|---|---|---|---|
| Ampère | 1823 | [17] | -2 | 3 | 0 | 0 | central force |
| Grassmann | 1845 | [11, 15] | -1 | 0 | 0 | 1 | no monopoles |
| Riemann | 1875 | [18] | -1 | 0 | 1 | 1 | moment conserved |
| Whittaker | 1912 | [1, 15] | see under Riemann | | | | |
| Brown | 1955 | ???? | 1 | -6 | 6 | 6 | ????? |
| Aspden | 1987 | [21] | -1 | 0 | 1 | -1 | cons.angular moment |
| Marinov | 1993 | [22] | -1 | 0 | 0.5 | 0.5 | |
| Cavallieri | 1998 | [19][23] | see under Grassmann | | | | experiment |

Of course this force law was not convincing as well. For the basic idea of Riemann and Whittaker was used by many others who built their "own" force laws using other assumptions. The discussion is running until today, see[23] and tab.1.

Now, due of Biot-Savart, cf. **(18),** or due to Ampère's observation, the force of the field **H** of a closed circuit on a differential current element is

$$d\mathbf{F} = \frac{i}{c} \mathbf{H} \times d\mathbf{s} \qquad (22)$$

If the influence of all field generating currents is summed up to the field **H** all possible field laws coincided to one field **H**

$$\mathbf{H} = \frac{I}{c} \oint \frac{\mathbf{x} \times d\mathbf{s}'}{|x^3|} \quad \text{or generally} \quad \mathbf{H} = \frac{1}{c} \oint \mathbf{j}(\mathbf{x}') \times \frac{\mathbf{x} - \mathbf{x}'}{|\mathbf{x} - \mathbf{x}'|^3} d^3 x' \qquad (23)$$

general scheme of proof for every magnetostatic force law:

According to a general theorem of vector analysis, see appendix 2, every vector field can be decomposed into a vortex field and a potential field. The vortex field is caused by currents, the potential field by charges. If this is compared with theorem 2 in appendix 2, then the Biot-Savart law generates a magnetic vortex field **H** of a current element. All other fields deviating from Biot-Savart, have to be written as

    field law = Biot-Savart-law + additional terms

These additional terms must be identified as a potential field. If the current is integrated over a closed circle the potential terms cancel to zero[1] .                                                                 □

If one integrates over both interacting closed circuits the von Neumann force law is obtained [1, 8, 24, 25]

$$\mathbf{F} = \frac{I_1 I_2}{c^2} \oint \oint \frac{\mathbf{x}_{12}}{|\mathbf{x}_{12}|^3} d\mathbf{s}_1 d\mathbf{s}_2 \qquad (24)$$

From **(22)** also follows, that the magnetic field can be calculated from a vector potential $\mathbf{H} := \nabla \times \mathbf{A}$ with

$$\mathbf{A}(\mathbf{x}) = \frac{1}{c} \oint \frac{\mathbf{j}(\mathbf{x}')}{|\mathbf{x}-\mathbf{x}'|} d^3x' \qquad (25)$$

---

[1] The Biot-Savart law is probably the correct version for physical currents. It does not generate "magnetic charges", it coincides with the **B**-field of a moving charge[19] according to Lienard-Wiechert(in the special case of zero acceleration) and takes into account of the self-interaction of single current elements, see Cavallieri et al.[19]. For measurements, see [23].

Then follows

$$\operatorname{div} \operatorname{rot} \mathbf{A} = \operatorname{div} \mathbf{H} = 0 \qquad (26)$$

So Ampère concluded: The cause of the magnetic field are not magnetic charges but only currents.

Ampère's theory includes as well para-, dia- oder ferromagnetic "excited" materials. The total magnetic field **B** includes the field from the measurable currents **j** and the field **M** of the magnetism of the material, where the field **M** (according to Ampère) is generated exclusively by currents in the material. Then follows

$$\operatorname{rot} \mathbf{B} = \frac{4\pi}{c}(\mathbf{j}_{\text{conductor}} + \mathbf{j}_{\text{material}}) \qquad (27)$$

with

$$\mathbf{B} := \mu \mathbf{H} \quad \text{oder} \quad \mathbf{B} := \mathbf{H} + 4\pi \mathbf{M} \qquad (28)$$

Analogously like for charges a relation is sought between the empirical variables. So the unknown current $\mathbf{j}_{\text{material}}$ is eliminated. If compared with electrostatics, see eq. **(14)** to **(17)**, it can be derived for currents (instead for charges), that for magnetostatics holds $\int \mathbf{j}\, d^3\mathbf{x} = \int \nabla . \mathbf{j}\, d^2\mathbf{x} = -\int \dot{\rho}\, d^2\mathbf{x} = 0$. Here is applied $\nabla . \mathbf{j} + \dot{\rho} = 0$ and $\mathbf{j}(\infty) = 0$, meaning that no currents exist at the boundary in the infinite. Thus, no charges can built up and only dipol terms and terms of higher orders of a series expansion of **B** can contribute to the field. Hence a definition **(29)** analogous to **(17)** can be used for the magnetization **M** of the material

$$\nabla \times \mathbf{M} := \mathbf{j}_{\text{material}} / c \tag{29}$$

Then, using **(27),(28)** and subtracting **(29)** Ampère´s laws are

$$\text{rot}\mathbf{H} = \frac{4\pi}{c} \mathbf{j}_{\text{conductor}} \qquad \text{div}\mathbf{B} = 0 \tag{30}$$

In order to derive the present version Ampère´s law is rewritten as [8]:

$$\nabla \times \mathbf{H} = \text{rot rot}\mathbf{A} = \text{grad div}\mathbf{A} - \nabla^2 \mathbf{A}$$
$$= \nabla \int \frac{\mathbf{j}(\mathbf{x}')}{c} \nabla \cdot \left( \frac{1}{|\mathbf{x}-\mathbf{x}'|} \right) d^3 x' - \int \frac{\mathbf{j}(\mathbf{x}')}{c} \nabla^2 \left( \frac{1}{|\mathbf{x}-\mathbf{x}'|} \right) d^3 x' \tag{31}$$

With the mathematical relations $\nabla |\mathbf{x}-\mathbf{x}'|^{-1} = -\nabla' |\mathbf{x}-\mathbf{x}'|^{-1}$ and $\nabla'^2 |\mathbf{x}-\mathbf{x}'|^{-1} = -4\pi \delta(\mathbf{x}-\mathbf{x}')$ this becomes

$$\nabla \times \mathbf{H} = -\nabla \int \frac{\mathbf{j}(\mathbf{x}')}{c} \cdot \nabla' \left( \frac{1}{|\mathbf{x}-\mathbf{x}'|} \right) d^3 x' + \frac{4\pi}{c} \mathbf{j} \tag{32}$$

because **A** also fulfills the Poisson equation $\nabla^2 \mathbf{A} = -4\pi \mathbf{j}/c$.

If the integral in **(32)** is integrated partially using that **j** vanishes at boundary in the infinite, then follows

$$\nabla \times \mathbf{H} = \frac{4\pi \mathbf{j}}{c} - \nabla \int \frac{\nabla' \mathbf{j}(\mathbf{x}')}{c|\mathbf{x}-\mathbf{x}'|} d^3 x' \tag{33}$$

Now, the observation is used that no charges build up during magnetostatic experiments. Using the continuity equation this fact can be translated into mathematics by $\nabla \cdot \mathbf{j} \sim -\dot{\varrho} = 0$. This yields Ampère´s law of magnetostatics:

$$\nabla \times \mathbf{H} = \frac{4\pi}{c}\mathbf{j} \qquad \text{or} \qquad \oint \mathbf{H}d\mathbf{s} = \frac{4\pi}{c}\int_S \mathbf{j}d\mathbf{A} \qquad (34)$$

Comparing the coefficients of **(31)** and **(33)** follows grad div **A**=0. Oftenly, it is assumed div **A** =0. This expression is known as the Coulomb-gauge. The vector potential **A** is not a unique function, because replacing **A** by **A**$^*$= **A** + $\nabla$f(**x**) fulfills the gauge as well. As we will see later it is necessary for the choice of the vector potential **A** that a physically motivated constraint has to be fulfilled for grad div **A**=**?** - i.e. the continuity equation [26]. At the time of Biot&Savart and Ampère this was not known fully and only the closed circuits could be tested out. So the result **(22)** for the **H**-Feld was ok. . However later, after the discovery of the electron by J.J. Thomson [27], discussions came up due to the basic problem behind the approaches of Biot&Savart and Ampère: Not every magnetic problem could be discussed by a closed electric circuit. Freely moving charges (as differential current elements) could exist and the question for their field had to be solved. So observations were published that longitudinal forces existed in railguns [28, 29] and in plasma tubes [30, 31] (See also the review article [32]). These forces seemed to be explained by Ampère's differential force law, but not by Biot-Savart´s version. Althought these problems seem to be solved today not in favour for longitunal forces[2] the problem will be left open here for further considerations. So all mathematically possible field configurations will be included in the discussion by adding a magnetic potential to the magnetic vector field. Mathematically any vector field **F** can be decomposed into two terms $\mathbf{F_C}$ and $\mathbf{F_V}$, derived from a potential (for $\mathbf{F_C}$) and from a vector potential of a vortex field (for $\mathbf{F_V}$), see the proof in appendix 2 [26] and [35, 36]. So any **H**-Feld can be described by

$$\mathbf{H} = \mathbf{H_V} + \mathbf{H_C} = \frac{1}{c} \int \mathbf{j}(\mathbf{x}') \times \frac{\mathbf{x}-\mathbf{x}'}{|\mathbf{x}-\mathbf{x}'|^3} d^3x' - \int \varrho_H(\mathbf{x}') \frac{\mathbf{x}-\mathbf{x}'}{|\mathbf{x}-\mathbf{x}'|^3} d^3x' \qquad (35)$$

Here $\varrho_H$ is the magnetic charge distribution due to the deviation from Biot-Savart´s differential law, see eq.**(18)**.

If a concrete system is solved with a boundary problem, a Laplace field $\mathbf{H_L}$ has to be added which fulfills rot $\mathbf{H_L}$ =0 and div $\mathbf{H_L}$ =0.

$$\mathbf{H} = \mathbf{H_V} + \mathbf{H_C} + \mathbf{H_L} = \frac{1}{c} \int \mathbf{j}(\mathbf{x}') \times \frac{\mathbf{x}-\mathbf{x}'}{|\mathbf{x}-\mathbf{x}'|^3} d^3x' - \int \varrho_H(\mathbf{x}') \frac{\mathbf{x}-\mathbf{x}'}{|\mathbf{x}-\mathbf{x}'|^3} d^3x' - \nabla\varphi(\mathbf{x}) \qquad (36)$$

Here is the Laplace field $\mathbf{H_L} := -\nabla\varphi(\mathbf{x})$. This potential describes a field, which is generated outside of the defined area of the problem. The field $\mathbf{H_L}$ helps to adapt the solution to the given

---

[2] Both observations were explained later by Rambaut &Vigier[33], see as well [34]. They pointed out, that these observations do not answer the question, because a closed moving circuit shows a "longitudinal" mechanical expansion due to a "expansion" pressure of a loop due to the Lorenz force.

boundary condition of the problem. Then **(36)** changes to

$$\mathbf{H} = \mathbf{H_V} + \mathbf{H_C} + \mathbf{H_L} = \nabla \times \mathbf{A} - \nabla \Xi - \nabla \varphi(\mathbf{x}) = \nabla \times \frac{1}{c} \int \frac{\mathbf{j}(\mathbf{x}')}{|\mathbf{x}-\mathbf{x}'|} d^3x' - \nabla \int \frac{\varrho_H(\mathbf{x}')}{|\mathbf{x}-\mathbf{x}'|} d^3x' - \nabla \varphi(\mathbf{x}) \quad \textbf{(37)}$$

with $\mathbf{A} := c^{-1} \int \mathbf{j}(\mathbf{x}')/|\mathbf{x}-\mathbf{x}'| d^3x'$ as magnetic vector potential and $\Xi := \int \varrho_H(\mathbf{x}')/|\mathbf{x}-\mathbf{x}'| d^3x'$ as potential function of the magnetic charges. If magnetic charges are included the magnetic field becomes a general field and loses all symmetry properties with respect of parity. So every field configuration can be described generally. It will be shown here that this is useful for problems with induced and permanent magnetization. Only a slight change in the conventional interpretation of the meaning of the magnetic field **B** leads to a Poisson equation for magnetic charges.

Proof:

The conventional theory assumes for problems with permanent magnetization [8] (without exciting field from outside), that

$$\nabla \cdot \mathbf{B_0} = \nabla \cdot (\mathbf{H_0} + 4\pi \mathbf{M}) = 0 \quad \textbf{(38)}$$

Here is **M** the magnetization of the material and $\mathbf{H_0}$ the inner magnetic field which generates the magnetization. If a field is applied additionally from outside, this equation is enlarged

$$\nabla \cdot \mathbf{B_0} = \nabla \cdot (\mathbf{H_0} + \mathbf{H} + 4\pi \mathbf{M}) = 0 \quad \textbf{(39)}$$

with **H** a the exciting magnetic field from outside, which is added. Because no currents are obvious in matter as cause for the inner $\mathbf{H_0}$-field it holds $\nabla \times \mathbf{H_0} = 0$. This means that $\mathbf{H_0}$ can be derived from a potential $\Xi_{\mathbf{H_0}}$ according to $\mathbf{H_0} = -\nabla \Xi_{\mathbf{H_0}}$ and the magnetostatic Poisson-equation follows [8]

$$\Delta \Xi_{\mathbf{H_0}} = -4\pi\rho_M \qquad (40)$$

with $\nabla.\mathbf{H_0} := 4\pi\rho_M$ defined as "effective magnetic charge density" in [8]. Our new definition of **B** is introduced here as follows

$$\nabla.\mathbf{B} := \nabla.(\mathbf{H} + 4\pi\mathbf{M}) = -4\pi\rho_M := 4\pi\rho_H \qquad (41)$$

This redefinition takes into account **(39)** and **(40)** and uses the redefinition $\rho_H := -\rho_M$. So the conventional equation **(39)** can be written down with magnetic charges in a form completely analogous to electrostatics. Analogously to electrostatics, the empirical field is now the **H**-field contrary to the conventional interpretation which takes the **B**-field. q.e.d. □

Thus, magnetostatic boundary problems can be worked out analogously to electrostatics with adapted boundary conditions. Textbooks show [8], that the solution of problems with induced magnetic polarization are completely identical to electrostatics. For the simple phenomena of induced polarization the outer boundary conditions represent either given current distributions of a coil exciting the material, either they represent an arbitrary field given on a chosen boundary. Then, the origin of this field at the boundary has to be thought in a distance far from the object under consideration.

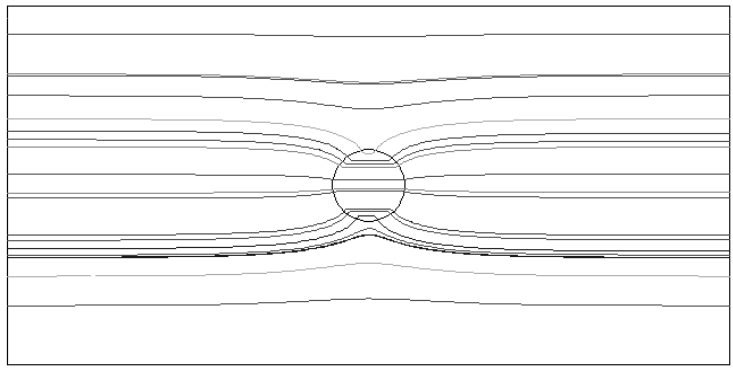

**fig.3: polarized bowl in a potential field**
boundary condition between inner and the outside of bowl: no charges and no currents, i.e. $\mathbf{B}_{inner} = \mathbf{B}_{outside}$ determines the charge distribution at the outer boundary of the bowl similar like in electro-statics. In the volumes holds the Laplace equation $\Delta\varphi = \nabla\mathbf{B} = 0$. For the equations of a metal bowl in the electric field the magnetic variables have to be replaced by electric ones analogously.

In or outside of the neighborhood of the induced magnetized body, however, the magnetic field of the material fulfills locally always $\Delta\varphi = \nabla.\mathbf{B} = 0$ and $\nabla\times\mathbf{H} = 0$. Here the Laplace equation holds for the induced magnetism and no charges exist ($\Delta\varphi=\nabla\mathbf{B}=\nabla\mathbf{D}=0$).

If the Laplace equation is not fulfilled then the existence of magnetic charges is probable. This can be the case for problems with permanent magnets. In this case the magnets are described by magnetic charges in their volume. The situation is still more complicated for the ferromagnetic hysteresis of iron. If compared with the conventional parity tabel, see tab.2, the **B**-field has (-1) parity under time inversion, i.e. if the current is inversed, the field has to be inversed as well. If a hysteresis exists, this is not the case, because the hysteresis line **B**(**H**) is not unique. For a change of parity with fields lower than the strenght of the coercivity, the change in parity can easily be disproved. In this case inhomogenities or gradients of magnetic permeability µ(**x**) can induce magnetic charges. Then it holds

$$\nabla.\mathbf{B}(\mathbf{x}) = \nabla.(\mu(\mathbf{x}).\mathbf{H}(\mathbf{x})) = \mu(\mathbf{x})\nabla.\mathbf{H}(\mathbf{x}) + \mathbf{H}(\mathbf{x}).\nabla\mu(\mathbf{x}) = 4\pi\varrho_M \neq 0 \qquad (42)$$

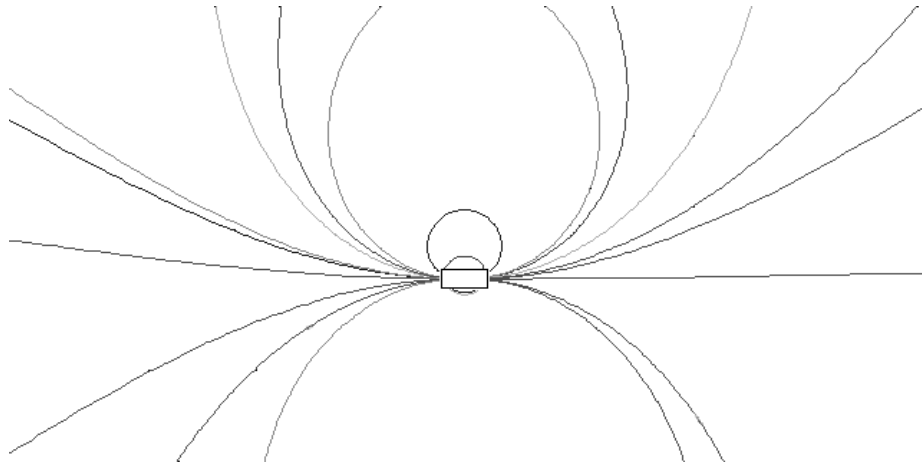

**fig.4: field lines of magnetic H-field of a cylindric permanent magnet**
the magnet is modelled here as capacity of magnetic charges. The magnetic charges are
distributed on the surfaces of north and south pole. The iron has a permeability of µ=10000

```
Tab.2: symmetry properties of conventional electrodynamics

It holds generally:   F(u) = P . F(-u)
```

| variable u | field F | parity P | sort of field |
|---|---|---|---|
| **x --> -x** | **E** | -1 | potential |
|  | **D** | -1 | potential |
|  | **H** | 1 | vortex |
|  | **B** | 1 | vortex |
|  |  |  | *cause* |
| t --> -t | **E** | 1 | charge |
|  | **D** | 1 | charge |
|  | **H** | -1 | current |
|  | **B** | -1 | current |

In fig.4 a magnet is modeled using magnetic charges. So, the presence of permanent magnetic poten-tial destroys the parity of a current-generated constant **B**-field similarly like it does the behaviour of the beta-decay in a field .

Hence, a general **B**-field with no parity cannot be explained solely by a vector potential **A**,

1) because **B** = rot **A** has always a defined parity;

2) because **B**:=rot **A** implies div rot **A** = div **B** = 0, which is contradicting to the physical result div **B**≠0.

So, the magnetic potential $\Xi$ has to be introduced for magnetic charges. Similarly like the magnetic vector potential **A** it has a more formal character, because it is not known very much about magnetic charges except of Ehrenhaft´s [5, 6] and Mikhailov´s experiments [37-48]. Important questions about concentrating, storing and conducting of magnetic charges are open.

The potentials of the magnetic field are

$$\mathbf{A}_{B,H,M} = \frac{1}{c}\int\frac{\mathbf{j}_{B,H,M}(\mathbf{x}')}{|\mathbf{x}-\mathbf{x}'|}d^3x' \qquad \Xi_{B,H,M} = \int\frac{\varrho_{B,H,M}(\mathbf{x}')}{|\mathbf{x}-\mathbf{x}'|}d^3x' \qquad (43)$$

Then, the magnetic fields can be derived

$$\mathbf{B}_V = \mathrm{rot}\mathbf{A}_B, \qquad \mathbf{B}_C = -\nabla\Xi_B$$

$$\mathbf{H}_V = \mathrm{rot}\mathbf{A}_H, \qquad \mathbf{H}_C = -\nabla\Xi_H \qquad (44)$$

$$\mathbf{M}_V = \mathrm{rot}\mathbf{A}_M/4\pi, \qquad \mathbf{M}_C = -\nabla\Xi_M/4\pi$$

using the definitions $\mathbf{B}_C := \mathbf{H}_C + 4\pi\mathbf{M}_C$, $\mathbf{B}_V := \mathbf{H}_V + 4\pi\mathbf{M}_V$, $\mathbf{A}_B := \mathbf{A}_H + \mathbf{A}_M$ and $\Xi_B := \Xi_H + \Xi_M$. The empirical magnetic field is

$$\mathbf{H} = \mathbf{H_V} + \mathbf{H_C} + \mathbf{H_L} \tag{45}$$

For the **B**-field holds:

$$\mathbf{B} := \mathbf{B_V} + \mathbf{B_C} + \mathbf{B_L} \tag{46}$$

Ampère´s laws are written (using rot $\mathbf{H_{C/L}}=0$ and div $\mathbf{B_{V/L}}=0$):

$$\nabla \times \mathbf{H} = \nabla \times \mathbf{H_V} = \frac{4\pi}{c}\mathbf{j} \qquad \nabla \cdot \mathbf{B} = \nabla \cdot \mathbf{B_C} = 4\pi\rho_B \tag{47}$$

The general force law of magnetism is then:

$$\mathbf{F} = \frac{1}{c}\int \mathbf{j} \times \mathbf{B_V}\, d\mathbf{x'}^3 + \int \varrho_H \cdot (\mathbf{H_C} + \mathbf{H_L})\, d\mathbf{x'}^3 \tag{48}$$

Later Ampère´s law $\nabla \times \mathbf{H_V} = 4\pi\mathbf{j}/c$ was extended by Maxwell. Maxwell realized [49], that this law could not describe cases, where electric charge appeared, which were stored up in capacitances. Maxwell solved the problem by a hypothesis, which turned out to be very useful, especially with respect to the theory of electromagnetic waves. He changed Ampère's equation to

$$\nabla \times \mathbf{H_V} = \frac{4\pi}{c}\mathbf{j} + \frac{1}{c}\frac{\partial \mathbf{D}}{\partial t} \tag{49}$$

Introducing the dielectric displacement d**D**/dt Maxwell removed a contradiction between physics and mathematics, because now the continuity equation could always be fulfilled as a constraint:

$$\operatorname{div}\operatorname{rot}\mathbf{H_V} = \operatorname{div}\left(\frac{4\pi}{c}\mathbf{j} + \frac{1}{c}\frac{d\mathbf{D}}{dt}\right) = \frac{4\pi}{c}\left(\operatorname{div}\mathbf{j} + \frac{d\varrho_E}{dt}\right) = 0 \tag{50}$$

This form of Ampère's law holds until today. It can describe as well the cases where charges are generated, for instance electron-positron pairs in high energy physics, electron-hole pairs in semiconductors, or dissociations into ions in chemistry. Maxwell's improvement does not change as well the gauge relation, because using **(31)** it can be calculated

$$\text{grad div}\mathbf{A} = -\nabla \int \frac{\nabla' \cdot \mathbf{j}(\mathbf{x}')}{c|\mathbf{x}-\mathbf{x}'|} d^3 x' = -\nabla \int \frac{\dot{\varrho}_E(\mathbf{x}')}{c|\mathbf{x}-\mathbf{x}'|} d^3 x' = \frac{1}{c}\frac{d\mathbf{D}}{dt} \qquad (51)$$

So the vector potential for Ampère's law **(34)** can be retained.

c) Faraday's law

The induction law has been found by Faraday. Using his formulation it is written

$$U = -\frac{d\Psi}{dt} \qquad (52)$$

For Faraday the flux $\Psi = \int \mathbf{B}\, d\mathbf{A}$ were the number of field lines, which go through a closed circuit. For an expanding or contracting circuit this is written today [9]

$$-\oint_C \mathbf{E} d\mathbf{s} = \frac{1}{c}\frac{d}{dt}\int_S \mathbf{B}\, d\mathbf{A} = \frac{1}{c}\int_S \frac{\partial \mathbf{B}}{\partial t} d\mathbf{A} + \frac{1}{c}\int_S \nabla\times(\mathbf{B}\times\mathbf{v})\, d\mathbf{A} + \frac{1}{c}\int_S \mathbf{v}(\nabla \cdot \mathbf{B})\, d\mathbf{A} \qquad (53)$$

A simple derivation can be done using the formalism of special relativity, see section e). This law can be formulated alternatively using **(53)**, $4\pi\varrho_H = \nabla \cdot \mathbf{B}$ and $\mathbf{j_H} = \rho_H \mathbf{v}$

$$-\nabla\times\mathbf{E} = \frac{1}{c}\frac{\partial\mathbf{B}}{\partial t} - \frac{1}{c}\nabla\times(\mathbf{v}\times\mathbf{B}) + \frac{4\pi}{c}\mathbf{j}_H \qquad (54)$$

It will be shown in the next section, that this equation is consistent with a gauge by a continuity equation for magnetic monopoles.

d) the complete Maxwell equations

The Maxwell equation describe the coupling of fields with moving charges in space. They can be generalized that they hold for solids and for gases and liquids.

The notations for the indices here are C:=charge, V:=vortex, E:=electric field, H:=magnetic field and g:=magnetic, e:=electric.
If magnetic charges are included the Maxwell equations are (using the definitions **v**=velocity and **j**:=**v**ρ)

$$-\oint \mathbf{E_V} d\mathbf{s} = \frac{d\Psi}{dt} := \frac{1}{c}\frac{d}{dt}\int \mathbf{B} d\mathbf{A} \qquad \oint \mathbf{H_V} d\mathbf{s} = \frac{d\Theta}{dt} := \frac{1}{c}\frac{d}{dt}\int \mathbf{D} d\mathbf{A} \quad \text{or}$$

$$-\nabla\times\mathbf{E_V} = \frac{1}{c}\frac{\partial\mathbf{B}}{\partial t} - \nabla\times(\frac{\mathbf{v_e}}{\mathbf{c}}\times\mathbf{B}) + \frac{4\pi}{c}\rho_H \mathbf{v_e} \qquad \nabla\times\mathbf{H_V} = \frac{1}{c}\frac{\partial\mathbf{D}}{\partial t} - \nabla\times(\frac{\mathbf{v_g}}{\mathbf{c}}\times\mathbf{D}) + \frac{4\pi}{c}\rho_E \mathbf{v_g}$$

$$\text{div } \mathbf{D_C} = 4\pi\varrho_E \qquad \qquad \text{div } \mathbf{B_C} = 4\pi\varrho_H \qquad (55)$$

$$\dot{\varrho}_E + \nabla\cdot\mathbf{j_E} = \text{div }\nabla\times(\frac{\mathbf{v_e}}{\mathbf{c}}\times\mathbf{B}) = 0 \qquad \dot{\varrho}_H + \nabla\cdot\mathbf{j_H} = \text{div }\nabla\times(\frac{\mathbf{v_g}}{\mathbf{c}}\times\mathbf{D}) = 0$$

$$\Delta\varphi_D = \text{div } \mathbf{D_L} = 0 \qquad \qquad \Delta\varphi_B = \text{div } \mathbf{B_L} = 0$$

$$\mathbf{D} = \mathbf{D_V} + \mathbf{D_C} + \mathbf{D_L} \qquad \qquad \mathbf{B} = \mathbf{B_V} + \mathbf{B_C} + \mathbf{B_L}$$

For a mixed system of charged particles the individual equations of each sort of particle have to be added together.
In the version above Ampère´s law is extended by the so called

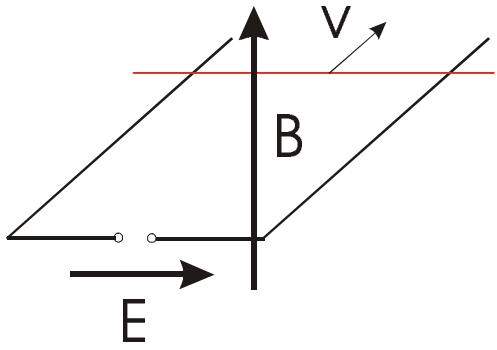 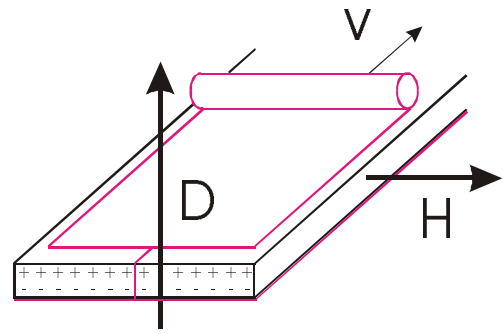

**fig.5a: E-field due to the Lorenz force**
at the expansion (or contraction) of a circuit in a magnetic field

**fig.5b: H-field due to the Rowlands force**
at the roll out of a conducting foil over a polarized electret material

Rowlands term which is electric analog to the Lorenz force. This term takes account for a **H**-field, which is generated, if a capacitance grows in an electric field, see fig.5b.
Similarly the Laplace field is accounted for in **(55)**.
Comparing with quantum mechanics we note, that in the first line of **(55)** the magnetic and the electric fluxes are quantized by the number n. It holds $\Psi = n.\Psi_0 = -4\pi.n.g$, where $\Psi_0 = -h.c/e$ is the elementary flux unit and g is the Dirac monopole, cf. first eq. **(55)** [74]. An analogous relation $\Theta_0 = -4\pi.e$ is valid for the electric charge unit e. Furthermore, the continuity equation for magnetic charges, cf. the third eq. **(55)**, is as well a continuity equation of angular moment, cf.[75]. Due to the conservation of angular momentum it should follow that such field structures can exist only if their permanent magnetic torque is compensated by an elastic or inertial torque of the body (cf. p.57).
Thus, a magnetic charge is a collective state of a spin field. To complete the electromagnetic theory an electric vector potential must also be introduced. It is generated by magnetic currents. All generating potentials are listed in **(56)**

$$\Phi_{D,E,P} = \int \frac{\varrho_{D,E,P}(\mathbf{x}')}{|\mathbf{x}-\mathbf{x}'|} d^3x' \qquad \Gamma_{D,E,P} = \frac{1}{c}\int \frac{\mathbf{j}_{B,H,M}(\mathbf{x}')}{|\mathbf{x}-\mathbf{x}'|} d^3x'$$

$$\mathbf{A}_{B,H,M} = \frac{1}{c}\int \frac{\mathbf{j}_{D,E,P}(\mathbf{x}')}{|\mathbf{x}-\mathbf{x}'|} d^3x' \qquad \Xi_{B,H,M} = \int \frac{\varrho_{B,H,M}(\mathbf{x}')}{|\mathbf{x}-\mathbf{x}'|} d^3x'$$

(56)

They are interconnected with the fields by

$$\begin{aligned}
\mathbf{B}_V &= \text{rot}\,\mathbf{A}_B, & \mathbf{B}_C &= -\nabla\Xi_B \\
\mathbf{H}_V &= \text{rot}\,\mathbf{A}_H, & \mathbf{H}_C &= -\nabla\Xi_H \\
\mathbf{M}_V &= \text{rot}\,\mathbf{A}_M/4\pi, & \mathbf{M}_C &= -\nabla\Xi_M/4\pi \\
\mathbf{D}_V &= \text{rot}\,\Gamma_D, & \mathbf{D}_C &= -\nabla\Phi_D \\
\mathbf{E}_V &= \text{rot}\,\Gamma_E, & \mathbf{E}_C &= -\nabla\Phi_E \\
\mathbf{P}_V &= \text{rot}\,\Gamma_P/4\pi, & \mathbf{P}_C &= -\nabla\Phi_P/4\pi
\end{aligned}$$

(57)

Summarizing it can be said about the Maxwell equations:
Electric and magnetic fields can be described mathematically as general fields. Their causes are charges and currents of electric and magnetic particles, which fulfill the continuity equation as a constraint. Due to the mathematics the electric and magnetic fields can be decomposed into a vortex, a potential field and a Laplace field. The charges build up the potential fields, the currents the vortex field and the Laplace field adapts to the boundary conditions.

e) The Maxwell equations and the theory of relativity
In the theory of relativity the Maxwell equations are formulated in the terminology of tensor calculus.

The theory of relativity relates the variables measured in a reference system to the variables of another system which moves relative to the first system. The transformation applies for a movement in z-direction (using the definitions $\beta := v/c, \gamma := 1/\sqrt{1-\beta^2}$)

$$a_{ij} = \frac{\partial x'_i}{\partial x_j} = \begin{pmatrix} \gamma & 0 & 0 & i\beta\gamma \\ 0 & 1 & 0 & 0 \\ 0 & 0 & 1 & 0 \\ -i\beta\gamma & 0 & 0 & \gamma \end{pmatrix} \qquad (58)$$

Similarly vectors are transformed (using the Einstein convention)

$$A'_i = a_{ij}.A_j \qquad (59)$$

Tensors $T'_{ij}$ are transformed by

$$T'_{ij} = a_{ik}.a_{jl}.T_{kl} \qquad (60)$$

The 4-vectors of the theory of relativity are, cf. appendix 3,

$$\begin{aligned}
\textit{space coordinates}: &\quad \mathbf{x} = (x, y, z, ict) \\
\textit{momentum}: &\quad \mathbf{p} = (p_x, p_y, p_z, imc) \\
\textit{wave number}: &\quad \mathbf{k} = (k_x, k_y, k_z, \frac{i}{c}\omega) \\
\textit{electric 4-current}: &\quad \mathbf{j_E} = (j_x^E, j_y^E, j_z^E, ic\rho^E) \\
\textit{magnetic 4-current}: &\quad \mathbf{j_H} = (j_x^H, j_y^H, j_z^H, ic\rho^H) \\
\textit{electric Lorenz vector}: &\quad \mathbf{L_E} = (A_x^E, A_y^E, A_z^E, ic\Phi^D) \\
\textit{magnetic Lorenz vector}: &\quad \mathbf{L_H} = (\Gamma_x^H, \Gamma_y^H, \Gamma_z^H, ic\Xi^B)
\end{aligned} \qquad (61)$$

The 4-vectors are invariant, i.e. the length of a vector is independent from the state of movement of the reference system. From this property and from **(61)** follows the continuity equation

$$\frac{d}{dx_i} j^i{}_E = \text{div}\, \mathbf{j_E} + \frac{d\varrho_E}{dt} = 0 \qquad \frac{d}{dx_i} j^i{}_H = \text{div}\, \mathbf{j_H} + \frac{d\varrho_H}{dt} = 0 \qquad (62)$$

An analogous equation - the Lorenz gauge - holds as well for Lorenz vectors, see appendix 3. The definitions for the electromagnetic tensor field at no current (**v**=0) are

$$F^{ij} := \begin{pmatrix} 0 & -E_3 & E_2 & -iB_1 \\ E_3 & 0 & -E_1 & -iB_2 \\ -E_2 & E_1 & 0 & -iB_3 \\ iB_1 & iB_2 & iB_3 & 0 \end{pmatrix} \qquad G^{ij} := \begin{pmatrix} 0 & H_3 & -H_2 & -iD_1 \\ -H_3 & 0 & H_1 & -iD_2 \\ H_2 & -H_1 & 0 & -iD_3 \\ iD_1 & iD_2 & iD_3 & 0 \end{pmatrix} \qquad (63)$$

If there no current is flowing, ($\mathbf{j}=\rho\mathbf{v}=0$) the 4-currents are

$$\mathbf{j_E} = (0,0,0, ic\rho^E) \qquad \mathbf{j_H} = (0,0,0, ic\rho^H) \qquad (64)$$

Then the Maxwell equations can be written

$$\frac{d}{dx_j} F^{ij} = 4\pi j_H{}^i \qquad \frac{d}{dx_j} G^{ij} = 4\pi j_E{}^i \qquad (65)$$

The complete system **(55)** of Maxwell equations follows if the charges move. This is described by the following coordinate transformation

$$\frac{d}{dx'_n} F'^{kn} = \frac{d}{dx'_n} \frac{\partial x'_n}{\partial x_j} \frac{\partial x'_k}{\partial x_i} F^{ij} = 4\pi \frac{\partial x'_k}{\partial x_i} j_H{}^i = 4\pi j'_H{}^k$$

$$\frac{d}{dx'_n} G'^{kn} = \frac{d}{dx'_n} \frac{\partial x'_n}{\partial x_j} \frac{\partial x'_k}{\partial x_i} G^{ij} = 4\pi \frac{\partial x'_k}{\partial x_i} j_E{}^i = 4\pi j'_E{}^k \qquad (66)$$

**(66)** represents the complete Maxwell equations in tensor notation, cf. **(55)**. One consequence should be emphasized:

If currents exist the complete Maxwell equations have to be applied including the terms of Lorenz and Rowlands force.

f) the electromagnetic tensors of momentum and energy

*The electromagnetic conservation of energy*

The power of an electrically and magnetically charged particle is (using $\mathbf{F_H} = q_H(\mathbf{H} - \frac{\mathbf{v}}{c} \times \mathbf{D})$, $\mathbf{F_E} = q_E(\mathbf{E} + \frac{\mathbf{v}}{c} \times \mathbf{B})$ and $\mathbf{F} = \mathbf{F_E} + \mathbf{F_H}$ )

$$\frac{dE_{mech}}{dt} = \mathbf{F}.\mathbf{v} := q_E.\mathbf{E}.\mathbf{v} + q_H.\mathbf{H}.\mathbf{v} \qquad (67)$$

This equation integrated over the whole space yields with $\mathbf{j} := \mathbf{v}\rho$

$$\frac{dE_{mech}}{dt} = \mathbf{F}.\mathbf{v} := \int (\mathbf{j_E}.\mathbf{E} + \mathbf{j_H}.\mathbf{H}) \, d\mathbf{x}^3 \qquad (68)$$

If the Maxwell equations are solved for the currents, (i.e. $\mathbf{j_H} = \frac{c}{4\pi}\left[-\nabla \times \mathbf{E} - \frac{1}{c}\frac{\partial \mathbf{B}}{\partial t} + \nabla \times (\frac{\mathbf{v}}{c} \times \mathbf{B})\right]$ and $\mathbf{j_E} = \frac{c}{4\pi}\left[\nabla \times \mathbf{H} - \frac{1}{c}\frac{\partial \mathbf{D}}{\partial t} + \nabla \times (\frac{\mathbf{v}}{c} \times \mathbf{D})\right]$ ) and inserted in **(67)**, and using $\nabla.(\mathbf{a} \times \mathbf{b}) = \mathbf{b}.(\nabla \times \mathbf{a}) - \mathbf{a}.(\nabla \times \mathbf{b})$, it follows a modified Poynting energy conservation equation for the energy density:

$$\frac{de_{mech}}{dt} = -\nabla.\mathbf{S} - \frac{\partial U}{\partial t} + \frac{c}{4\pi}\left(\nabla \times (\frac{\mathbf{v}}{c} \times \mathbf{B})\right).\mathbf{H} + \frac{c}{4\pi}\left(\nabla \times (\frac{\mathbf{v}}{c} \times \mathbf{D})\right).\mathbf{E} \qquad (69)$$

Here the following definitions have been used

$$S := \frac{c}{4\pi}(\mathbf{E} \times \mathbf{H})$$

$$\frac{dU}{dt} := \frac{1}{4\pi}\left(\mathbf{E}\frac{d\mathbf{D}}{dt} + \mathbf{H}\frac{d\mathbf{B}}{dt}\right) \qquad (70)$$

The last two terms in **(69)** are non-standard, because the energy conservation is derived always without Rowlands and Lorenz terms.

*The electromagnetic conservation of momentum*

The force on a charge distribution of electromagnetic charge is

$$F_{mech} = \int \rho_E (\mathbf{E} + \frac{\mathbf{j_E}}{c} \times \mathbf{B}) + \rho_H (\mathbf{H} + \frac{\mathbf{j_H}}{c} \times \mathbf{D})\, d\mathbf{x'}^3 \qquad (71)$$

Using again Maxwell's equations solved for **j** this can be written

$$F_{mech} = \int [\mathbf{E}\nabla\cdot\mathbf{D} + \mathbf{H}\nabla\cdot\mathbf{B} + (\nabla\times\mathbf{H}^*)\times\mathbf{B} + (\nabla\times\mathbf{E}^*)\times\mathbf{D} - \frac{1}{c}\left(\frac{\partial\mathbf{D}}{\partial t}\times\mathbf{B}\right) + \frac{1}{c}\left(\frac{\partial\mathbf{B}}{\partial t}\times\mathbf{D}\right)]d\mathbf{x'}^3 \qquad (72)$$

From the footnote[3] and the definitions $\mathbf{E}^* := \mathbf{E} + \mathbf{v}/c \times \mathbf{B}$ and

$$\mathbf{E}\cdot\nabla\mathbf{D} + \mathbf{H}\cdot\nabla\mathbf{B} + (\nabla\times\mathbf{H})\times\mathbf{B} + (\nabla\times\mathbf{E})\times\mathbf{D} = \varepsilon_{ijk}\varepsilon_{jls}\frac{\partial E_s}{\partial x_l}D_k + \varepsilon_{ijk}\varepsilon_{jls}\frac{\partial H_s}{\partial x_l}B_k.$$

---

[3] The first three vector terms can be written in the terminology of the tensor calculus:

Using $\varepsilon_{ijk}\varepsilon_{jls} = \delta_{kl}\delta_{is} - \delta_{ks}\delta_{il}$ the first term is transformed to

$$\varepsilon_{ijk}\varepsilon_{jls}\frac{\partial E_s}{\partial x_l}D_k + \varepsilon_{ijk}\varepsilon_{jls}\frac{\partial H_s}{\partial x_l}B_k = E_i\frac{\partial D_j}{\partial x_j} + D_i\frac{\partial E_k}{\partial x_k} - D_k\frac{\partial E_k}{\partial x_i} + H_i\frac{\partial B_j}{\partial x_j} + B_i\frac{\partial H_k}{\partial x_k} - B_k\frac{\partial H_k}{\partial x_i}$$

In the 2$^{nd}$ and 5$^{th}$ term k can be exchanged with j without changing the result. The first three terms of **(73)** then follow.

$\mathbf{H}^* := \mathbf{H} - \mathbf{v}/c \times \mathbf{D}$ follows the balance

$$F_{mech} = \int \left[ \frac{d}{dx_k} \mathbf{T}_{ik}^* - \mathbf{D} \frac{d\mathbf{E}}{dx_k} - \mathbf{B} \frac{d\mathbf{H}}{dx_k} - \frac{d}{dt} \frac{\mathbf{D} \times \mathbf{B}}{4\pi c} \right. \\ \left. + \left( \nabla \times (\frac{\mathbf{v}}{c} \times \mathbf{B})) \times \mathbf{B} \right) + \left( \nabla \times (\frac{\mathbf{v}}{c} \times \mathbf{D})) \times \mathbf{D} \right) \right] d\mathbf{x'}^3 \quad (73)$$

Here $\mathbf{T}_{ik}^*$ is defined as $\mathbf{T}_{ik}^* := E_i D_j + H_i B_j$. The fourth term of the first line of **(73)** is $\mathbf{p}_{Feld} := (\mathbf{D} \times \mathbf{B})/(4\pi c)$ which is defined as the electromagnetic momentum $\mathbf{p}_{Feld}$ of the field.

If the generality of **(73)** is restricted (i.e. if only materials are used with purely linear constitutive relations like $\mathbf{B} = \mu \mathbf{H}$ and $\mathbf{D} = \varepsilon \mathbf{E}$) then the first three terms of **(73)** represent the Maxwell energy tensor:

$$\frac{d\mathbf{T}_{ik}}{dx_k} := \frac{d}{dx_k} (E_i D_k + H_i B_k - \frac{\delta_{ik}}{2} (\varepsilon \mathbf{E}^2 + \mu \mathbf{H}^2)) \quad (74)$$

This equation is found in the textbooks normally. The last two terms of **(73)** are omitted always, because "shorted" Maxwell equation are used which is wrong in the general case according to the author´s opinion.

The equations of conservation of energy and momentum describe the behaviour of a generalized capacitive-inductive- electronic element. Special cases for the energy equation are the pure capacitance (if $\mathbf{H} = 0$ and $\mathbf{B} = 0$) and the pure coil (if $\mathbf{E} = 0$ and $\mathbf{D} = 0$), see **(73)**. For these special cases the equation says, that the energy flowing into the electronic element can be identified with the electric or magnetic field energy.

Using **(69)** the definition of electromagnetic work is

$$W_{el} = \int \frac{dE_{mech}}{dt} \, dt \tag{75}$$

It should be said that the discussion about the "correct" equations **(69)** and **(73)** is alive until today, cf. [50].

It is remarkable that the derivation with monopoles yields the same result as without. The cause of this may be, that many Maxwell equations are the solutions from the theory of general relativity, because one degree of freedom remains undetermined during the derivation [8] [51, 52]. The considerations were done only for shorted Maxwell equations by the authors, i.e.

$$\begin{aligned} \nabla \cdot \mathbf{D} &= 4\pi\rho_E & \nabla \times \mathbf{H} &= \frac{1}{c}\frac{d\mathbf{B}}{dt} + \frac{4\pi}{c}\mathbf{j_H} \\ \nabla \cdot \mathbf{B} &= 4\pi\rho_H & -\nabla \times \mathbf{E} &= \frac{1}{c}\frac{d\mathbf{D}}{dt} + \frac{4\pi}{c}\mathbf{j_E} \end{aligned} \tag{76}$$

It can be shown, that all these equation can be transformed by

$$\begin{aligned} \mathbf{E} &= \mathbf{E}'\cos\zeta + \mathbf{H}'\sin\zeta & \mathbf{D} &= \mathbf{D}'\cos\zeta + \mathbf{B}'\sin\zeta \\ \mathbf{H} &= -\mathbf{E}'\sin\zeta + \mathbf{H}'\cos\zeta & \mathbf{B} &= -\mathbf{D}'\sin\zeta + \mathbf{B}'\cos\zeta \\ \rho_E &= \rho_E'\cos\zeta + \rho_H'\sin\zeta & \mathbf{j_E} &= \mathbf{j}_E'\cos\zeta + \mathbf{j}_H'\sin\zeta \\ \rho_H &= -\rho_E'\sin\zeta + \rho_H'\cos\zeta & \mathbf{j_H} &= -\mathbf{j}_E'\sin\zeta + \mathbf{j}_H'\cos\zeta \end{aligned} \tag{77}$$

If the parameter $\zeta$ in **(77)** is chosen appropriately, the conventional Maxwell equations without magnetic charges are the result. It is shown that relativistic pressure tensor (shorted calculation without Lorenz and Rowlands terms !) is invariant under these transformations.

If one believes, that every electric charge is in a constant proportion with a magnetic charge, -so the argumentation and the calculation of Harrison[53] and Katz [52]- the combined charge is regarded as a new "elementary charge", and built up a transformed (shorted) system of Maxwell equations with div **B**=0 [8, 52, 53]. So it is understandable, that Mikhailov [38, 48, 54] tried to determine the proportion between electric and magnetic charge, especially because the first workout of his measurements [38] spoke against the generally accepted theoretical value of Dirac [55, 56]. Anyway, in the light of these opinions of Harrison[53] and Katz [52], one can ask why Mikhailov sees any effects at all. For author the discussion is not at the end here. Perhaps, parity checks can solve this question.

g) boundary conditions

*stationary discontinous boundary conditions by charges*

In order to derive boundary condition equation **(13)** is applied on a fictive "pillbox" at the boundary between two materials of a potential field [8], see fig.6a .

So one obtains the relation (with σ:=surface charge density)

$$\int \nabla . \mathbf{F_C} \, dV = \int_S \mathbf{F_C} . \mathbf{n} \, da = (\mathbf{F_C(1)} - \mathbf{F_C(2)}) . \mathbf{n} \, \triangle a = 4\pi\sigma\triangle a \tag{78}$$

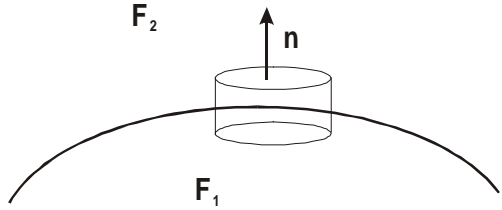 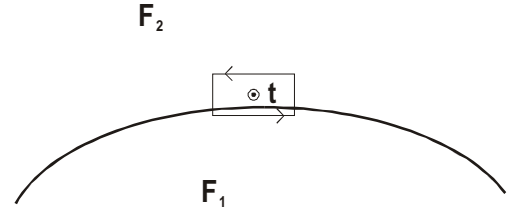

**fig.6a the pillbox - construction**
for the determination of boundary conditions due to charges

**fig.6b the circuit - construction**
for the determination of boundary conditions due to currents

Equation **(78)** shows a relation between vector components of the field **F₁** in region 1 and **F₂** in region 2 which both are normal to the surface. This yields for the vertical components of the dielectric displacement **D_C** :

$$(\mathbf{D_C}(1) - \mathbf{D_C}(2)) \cdot \vec{n} = 4\pi\sigma_E \tag{79}$$

i.e. at the boundary there is a discontinuity which is determined by the surface charge density. An analog holds for the vertical component of the magnetic field **B_C** :

$$(\mathbf{B_C}(1) - \mathbf{B_C}(2)) \cdot \vec{n} = 4\pi\sigma_H \tag{80}$$

For an electric or magnetic conducting surface holds

$$\Phi = constant \qquad \Xi = constant \tag{81}$$

*stationary boundary conditions by currents*

Equation **(29)** can be applied to derive a boundary condition if a surface current **k** flows at the boundary between regions of different materials, see fig.6b . So one obtains [8]

$$\int \nabla \times \mathbf{F_V} \, dA = \int_S \mathbf{F_V} \, d\mathbf{s} = (\mathbf{n} \times \mathbf{t}) \cdot (\mathbf{F_V}(1) - \mathbf{F_V}(2)) \Delta l = \frac{4\pi}{c} \mathbf{k} \cdot \mathbf{t} \, \Delta l \quad (82)$$

Equation **(82)** is a relation between the vector components **F₁** and **F₂** which flow tangentially on the surface of the boundary between two regions 1 and 2 of different materials.

*discontinuities of the magnetic vortex field*
for tangents to the surface

$$\vec{\mathbf{n}} \times (\mathbf{H_V}(1) - \mathbf{H_V}(2)) = \frac{4\pi}{c} \mathbf{K_E} \quad (83)$$

*discontinuity of the electric vortex field*
for tangents to the surface

$$\vec{\mathbf{n}} \times (\mathbf{E_V}(1) - \mathbf{E_V}(2)) = \frac{4\pi}{c} \mathbf{K_H} \quad (84)$$

For more general, nonstationary boundary conditions at moving surfaces, see [8].

h) the constitutive equations of the material

The system of Maxwell equations can be solved after the constitutive equation are known which describe the material properties. They couple the electric variables (**E,D**) and the magnetic variables (**B,H**) which can be represented generally by

$$\begin{pmatrix} \mathbf{D} \\ \mathbf{B} \end{pmatrix} = coupling \circ \begin{pmatrix} \mathbf{E} \\ \mathbf{H} \end{pmatrix} \tag{85}$$

In the most cases these couplings are simple, i.e.

$$\begin{aligned} as \quad resistor: & \quad \mathbf{j} = \sigma.\mathbf{E} \\ or \ capacitively: & \quad \mathbf{D} = \varepsilon\mathbf{E} \\ or \ inductively: & \quad \mathbf{B} = \mu\mathbf{H} \end{aligned} \tag{86}$$

Initially the material constant were constants which described the simple cases of material properties. Later more complicated nonlinear functions were found which could generate phase transitions, i.e.

$$\sigma = \sigma(\mathbf{E}), \quad \varepsilon = \varepsilon(\mathbf{E}), \quad \mu = \mu(\mathbf{H}) \tag{87}$$

After the fundamental crystal structures were known, the material properties could be correlated to the symmetry of the crystals. Then, the constitutive equation were described by tensors

$$\boldsymbol{\sigma} = \sigma_{ik}(\mathbf{E}), \quad \boldsymbol{\varepsilon} = \varepsilon_{ik}(\mathbf{E}), \quad \boldsymbol{\mu} = \mu_{ik}(\mathbf{H}) \tag{88}$$

which were first linear, then non-linear.
Then, materials were discovered whose properties were magnetic and electric, and where an electric field influenced the magnetic properties and vice versa [57] [58].
The theory of relativity found out that dielectric or magnetic polarized material behaved different if it was set in motion. The following equations are from [59]

$$\mathbf{E}' = \mathbf{E} - \frac{\mathbf{v}}{c} \times \mathbf{M}$$

$$\mathbf{H}' = \mathbf{H} - \frac{\mathbf{v}}{c} \times \mathbf{P} \tag{89}$$

A further complication of the constitutive relations are space-dependence of the material properties which are realized for instance as electronic elements.

Furthermore all materials have their own dynamics in time in the form of relaxation time.

If all material properties are accounted for then the general constitutive equations can be abstracted as additional differential equations which help to solve the complete system of partial differential equations. This system can be written as

$$\begin{pmatrix} \dot{\mathbf{E}} \\ \dot{\mathbf{H}} \end{pmatrix} = \begin{pmatrix} f_1(\mathbf{E},\mathbf{D},\mathbf{H},\mathbf{B};T,\rho_i,\dot{\mathbf{x}},\omega,...)(\mathbf{x},t) \\ f_2(\mathbf{E},\mathbf{D},\mathbf{H},\mathbf{B};T,\rho_i,\dot{\mathbf{x}},\omega,....)(\mathbf{x},t) \end{pmatrix} \tag{90}$$

or

$$\begin{pmatrix} \dot{\mathbf{D}} \\ \dot{\mathbf{B}} \end{pmatrix} = \begin{pmatrix} f_1(\mathbf{E},\mathbf{D},\mathbf{H},\mathbf{B};T,\rho_i,\dot{\mathbf{x}},\omega,...)(\mathbf{x},t) \\ f_2(\mathbf{E},\mathbf{D},\mathbf{H},\mathbf{B};T,\rho_i,\dot{\mathbf{x}},\omega,....)(\mathbf{x},t) \end{pmatrix} \tag{91}$$

The variables after the semicolon show that the constitutive equations may not depend only from electromagnetic parameters, but can depend as well from mechanic or thermodynamic material properties. This means that the electrodynamics cannot be separated from the other areas of physics. If these the material properties drift under the influence of electromagnetic fields then a purely electrodynamic description is not sufficient and further differential equations from other areas of physics have to be added to a complete partial differential equation system.

Examples:

1) Known examples are electric motors and generators. Here the mechanic equations of motion of the motors are added. They describe the motion by the angular coordinate of the rotor.

2) Other systems are magnetic materials, for which the Landau-Lifshitz-Gilbert - equation [57] [60] hold

$$\dot{\mathbf{M}} = -\gamma.\mathbf{M}\times\mathbf{H}_{\text{eff}} + \alpha.\mathbf{M}\times(\mathbf{M}\times\mathbf{H}_{\text{eff}}) \qquad (92)$$

It generates a system of partial differential equation (:=PDE) if it is combined with the equation of the magnetostatic potential **(41)** [61]. It allows to calculate magnetic domains in ferromagnetic materials.

3) A homogeneous thermostatic system like a polymer solution is described by a free energy density $f$. The system plus field is described by the free energy density $f^*=f+\rho_E*\Phi_E$. Then using the definitions of the global chemical potential $\mu_i^* := df^*/dx_i$ and $x_i :=$ volume ratio the PDE-system hold

$$\begin{aligned}&\Delta\mathbf{\Phi_D}(x_i(\mathrm{r}),\mathrm{r}) = -4\pi\rho_E \\ &\frac{\partial\mu_i^*}{\partial\mathrm{r}}(x_i(\mathrm{r}),\mathbf{\Phi_E}(\mathrm{r})) = 0\end{aligned} \qquad (93)$$

For a magnetic system (for instance a ferrofluid solution) the electric variables (**E**,**D**) are replaced by magnetic ones (**H**,**B**). The magnetic charge density ρ is set to zero, because no magnetic charges can be detected during the magnetization, see [62].

4) If the problem depends from time additionally, it is necessary to replace the second equation of **(93)** by the thermodynamic functions for non-equilibrium. Then one can write

$$\Delta \mathbf{\Phi_D}(x_i(r),r) = -4\pi\rho_E$$

$$j_i = -D_i \frac{\partial n_i(r)}{\partial r} + \lambda_i \frac{Z_i e n_i(r) \mathbf{E}(r)}{RT}$$

**(94)**

Here hold the definitions n:=concentration, r:=space coordinate Z:=number of charges per ion, e:=elementary charge, **E**:= electric field, **R:**=Avogadro-constant, T:=temperature, D:=diffusion constant, λ:=mobility. The second equation of **(94)** is the Nernst-Planck equation, which should coincide with the second equation **(93)** for **j**=0. So electrochemical problems are discussed, cf.[63].

5) In semiconductors the charge densities depend from chemical potential or quasi-Fermi level, which can be influenced by the electric potential. A good example for such a system is a InAs-quantum dot-doted FET invented by Yusa&Sakaki [64]. Its structure is shown in fig.7. The FET can be used for storing data by charging the gate capacitance.

The theoretical model of this FET stems from Rack et al.[65]. The PDE´s of the system is:

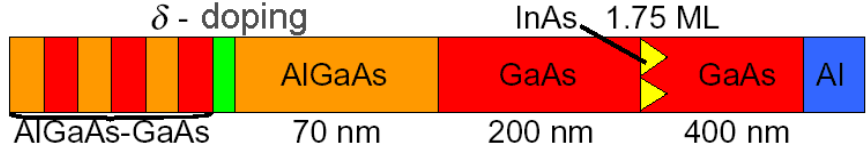

**fig.7: structure of a InAs-quantum dot-doted GaAs-FET**
a two-dimensional electron gas (2DEG) is located in the boundary between AlGaAs and GaAs. It represents the zero potential of the system. The electric potential is applied to the Al layer, cf. figs.9

$$\begin{aligned}
&\textit{Poisson-equation}: \quad \varepsilon_0 \partial_z[\varepsilon(z).\partial_z\Phi(z)] = -\rho(z) \text{ with } \rho(z) = e[N_D^+(z) - n^{3d}(z) - n_{QD}(z)] \\
&\textit{current}: \quad \partial_t n(z) = \frac{1}{e}\partial_z j(z) - f(n_{QD}(z,t), n(z)) = 0 \quad\quad (95) \\
&\textit{recombinations}: \quad \partial_t n_{QD}(z,t) = f(n_{QD}(z,t), n(z))
\end{aligned}$$

Here are $\varepsilon_0$ := dielectric constant of vacuum, $\varepsilon$ :=dielectric constant of the material, $\rho$:=charge density, $N_D$:=density of donators, $n_{3d}$:=charge density of electrons, $n_{QD}$:=charge density of electron trapped im quantum dots, $n(z)$:=free electron density function specified in the article, $j$:=current in the FET, and $f(n_{QD},n)$ is a specific function, which characterizes the recombination process, see [65]. Figs.8 show the electron density in the 2DEG versus voltage. Remarkable is the orientation of the electric cycle which is opposite to the ferroelectric loss hysteresis. This suggests a "gain hysteresis".

It is known that electric work can be changed to mechanic work with efficiencies until 100% in the best electromotors. So electric work should be equivalent to mechanical work in a thermodynamical sense. An "isothermically" proceeded electric

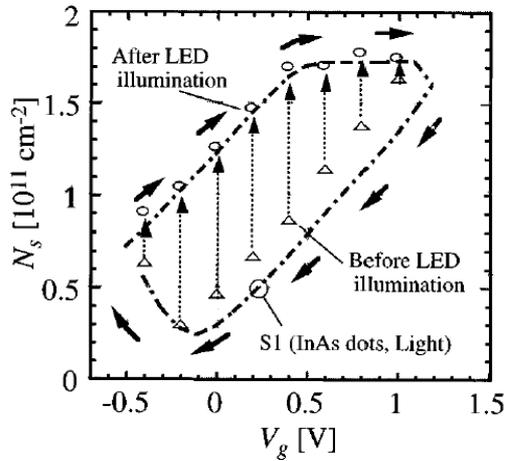 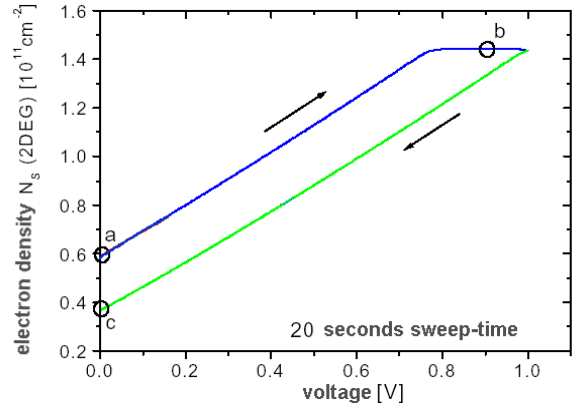

**fig.8a the experiment of Yusa-Sakaki- cf. [64] hysteresis of a InAs-quantum dot-doted FET**
electron charge density of the two-dimensional electron gas (2DEG) vs. gate voltage

**fig.8b the theoretical calculation of the Yusa-Sakaki-FET by Rack et al.**
electron charge density of the two-dimensional electron gas (2DEG) vs. gate voltage cycle with an orientation like in fig.8a can fulfill the energy balance only if heat flows in from outside. Thus, the FET is a candidate for second law violation because only heat and electricity can be exchanged. According to own recent work [62] such cycles could be possible and further evidence can be found: Cooling effects in semiconductors have been predicted by [66]. These considerations support the considerations for the FET discussed above. According to [66] the FET is cooled down if it is set under voltage. So the electrons are enforced into the quantum dots below the quasi-Fermi niveau, which leads to a cooling down because the state space for the electrons is enlarged adiabatically .

The whole cycle is completed as follows: The electrons remain sticking at the quantum dots due to their binding energy, even after the discharge of the FET. Then, the FET goes back to the equilibrium either if the voltage is slightly inverted, cf.fig.8a, either if the wavelength of the thermal radiation is suffiently high to overcome the binding energy of 0.25eV, which holds the electrons in the quantum dot potentials. So, the system can be regarded also as a concretisation of Maxwell´s demon. The electric energy is lended probably from the quantum dots to be paid back after some time from the thermic influx of environment. Further evidence for this idea can be found from the results of fig. 9a-c, which show the conduction band edge (which is here equivalent to the potential) in the FET at the beginning of the cycle, after charging it with voltage, and after discharging the capacitance. From the slope in the diagrams one calculates the electric fields in the FET. If one regards the FET as a capacitance and applies **(69)** one can estimate the energy exchanged after a cycle. From **(69)** follows for a pure capacitance

$$\Delta W = \int_0^T U.I\, dt = -\frac{1}{4\pi} \iint \mathbf{E}\, d\mathbf{D}\, dV \qquad (96)$$

If one reads off electric field values from the slopes in fig. 9a to fig. 9c one obtains the electric field energies in the FET:
before charging the gate capacitance
$\quad W_1 \sim \mathbf{E}^2 * V \sim (1V/600nm)^2 * 600nm = 0.00166666$
after discharging the gate capacitance
$W_2 \sim \sum \mathbf{E}^2_i * V_i \sim (.38V/200nm)^2 * 200nm + (.62V/400nm)^2 * 0.400nm = 0.001683$
energy balance: $\quad \underline{\Delta W \sim -(W_2-W_1) \sim -0.00001633}$

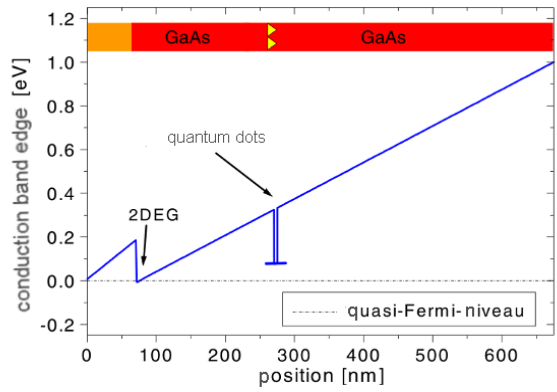

**fig.9a the conduction band edge vs. position in the FET of Yusa&Sakaki**
before the cycle: voltage U=0 V

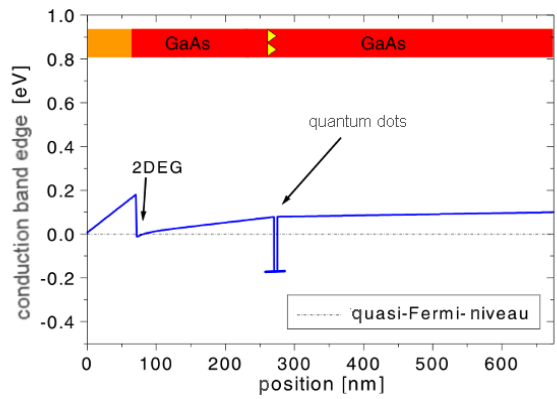

**fig.9b the conduction band edge vs. position in the FET of Yusa&Sakaki**
in the cycle: voltage U=0.9 V

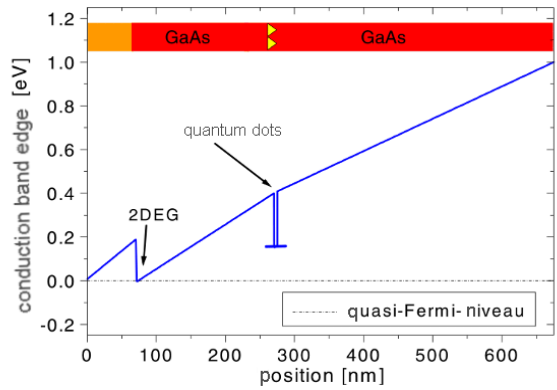

**fig.9c the conduction band edge vs. position in the FET of Yusa&Sakaki**
after the cycle: voltage U=0 V
the band edge is changed due to the storage of charges in the quantum dots, cf. fig.9a The energy difference of ~1% is negative meaning that electric energy is released by the FET after the electric cycle is closed. The Second Law is violated by the hysteresis of the equilibrium state. The effect is due to the nonlinear behaviour of the FET. Of course, all evidence of the experiment with the Yusa-Sakaki FET is indirectly concluded here. More decisive would be a full balance of all electrons in the calculation or the experiment. Herewith, constitutive equations are characterized from the simple case to the most complicated systems. Generally, the description of a system may be very sophisticated. However, normally the description is made as simple as possible.

## 3. Conclusions

It has been shown that the existence of magnetic charges is justified at least as a mathematical tool especially if fields of permanent magnetism have to be described. Physically these results suggest the following consequences to be proved:

If magnetic charges can be separated in space - for instance by the form of the distribution of polarisation in a permanent magnet - and if this magnet moves in a circle, two opposite magnetic currents are generated which itself should generate an electric field according to Faraday´s law extended for magnetic charge currents. Measurement of the electric field from moving permanent magnets can answer the question whether the electric field stems from changing magnetic fields or from moving magnetic charges. Both possibilities are calculable.

The setup of such experiments would be similar to constructions from the unofficial subscene of physics. J. Searl [67-70], D. Hamel [71] and Godin&Roschin [72, 73] claim to have observed strong electrostatic effects around moving permanent magnets. Fig.10 shows the calculated electrical vortex field Γ due to a moving permanent magnet ring representing two currents of opposite magnetic surface charge which are placed on top and bottom of the ring, cf. fig.4. The electric field is calculated from the electric vortex field by **E** = rot Γ, see fig.11a and fig. 11b: the pictures show the electric field strength and the position angle of the field around the cross section of the right half of the ring. Appendix 4 shows the method of the calculation. The order of magnitude coincides with Godin&Roschin [72, 73].

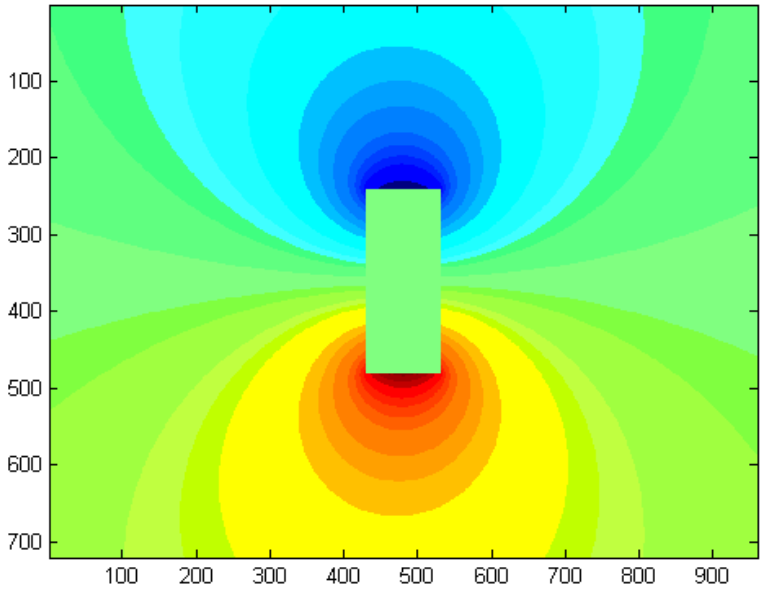

**fig.10: strength of the Φ-component of the electric vortex field Γ of a rotating magnetic ring**
cross section view: ring radius 1m, ring width 5cm, ring heigth 12cm, center of rotation is to the
left, (not to be seen in picture). rainbow scale: blue is minus min., red is plus max., see appendix 4.

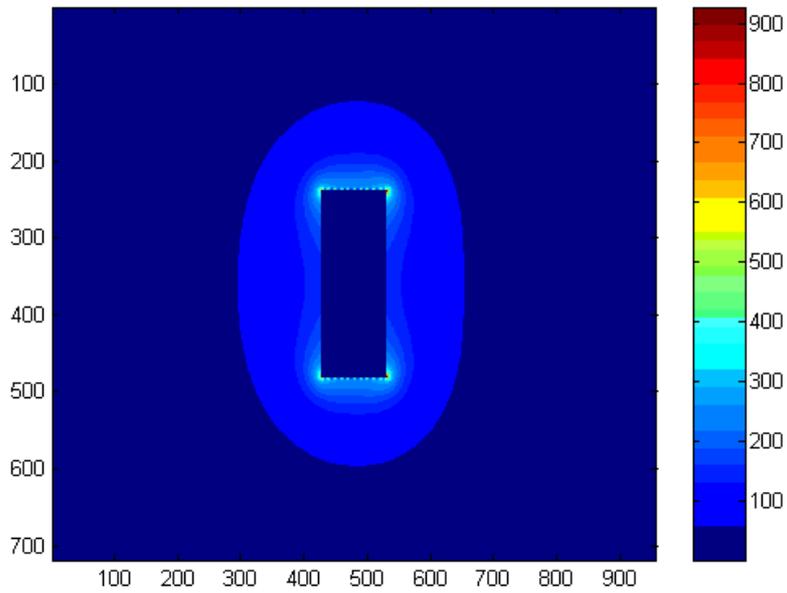

**fig.11a: E-field strength around a rotating magnetic ring (cross section)**
ring radius 1m, ring width 5cm, ring heigth 12cm, center of rotation is to the left, (not shown in the picture). Arbitrary units. Picture is calculated from the data of fig.10, see appendix 4.

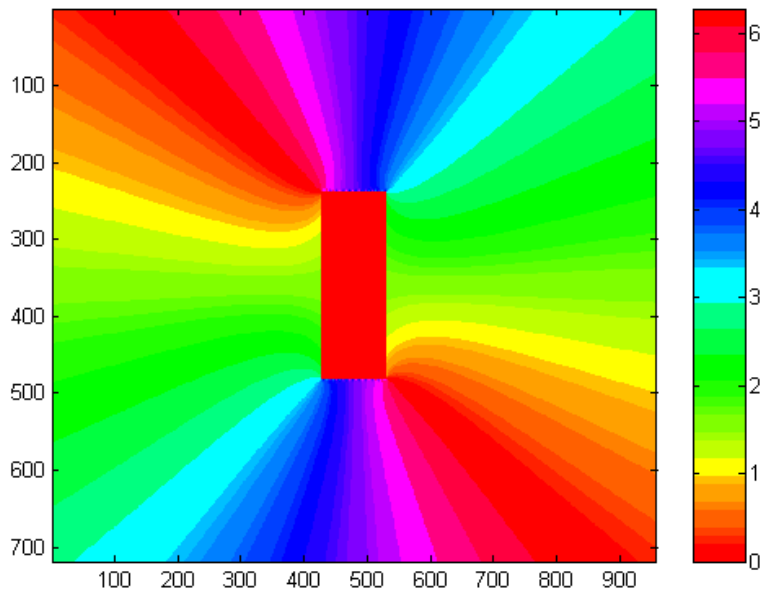

**fig.11b position angle of the E-field around a rotating magnetic ring (cross section)**
radius 1m,ring width 5cm,ring heigth 12cm,center of rotation is to the left,(not shown in the picture)

**Appendix 1**: the derivation of the multipole expansion

First the term $1/|\mathbf{x}-\mathbf{x}'|$ is written as:

$$\frac{1}{|\mathbf{x}-\mathbf{x}'|} = \frac{1}{\sqrt{\mathbf{x}^2+\mathbf{x}'^2-2\mathbf{x}.\mathbf{x}'}} = \frac{1}{|\mathbf{x}|}\frac{1}{\sqrt{1+\frac{\mathbf{x}'^2-2\mathbf{x}.\mathbf{x}'}{|\mathbf{x}|^2}}}$$

with the abbreviation $\alpha := (\mathbf{x}'^2-2\mathbf{x}.\mathbf{x}')/|\mathbf{x}|^2 \ll 1$.

This expression is expanded in a series

$$\frac{1}{\sqrt{1+\alpha}} = 1 - \frac{\alpha}{2} + \frac{3}{8}\alpha^2 \pm ..... = 1 - \frac{1}{2}\frac{\mathbf{x}'^2}{|\mathbf{x}|^2} + \frac{2}{2}\frac{\mathbf{x}.\mathbf{x}'}{|\mathbf{x}|^2} + \frac{3}{8}\left[\frac{\mathbf{x}'^2-2\mathbf{x}.\mathbf{x}'}{|\mathbf{x}|^2}\right]^2 \pm ....$$

Using the definitions $\mathbf{x}_0 := \mathbf{x}/|\mathbf{x}|$ and $|\mathbf{x}| := r$ one obtains

$$\frac{1}{|\mathbf{x}-\mathbf{x}'|} = \frac{1}{r} + \frac{1}{r^2}(\mathbf{x}'.\mathbf{x}_0) + \frac{1}{r^3}[\frac{3}{2}(\mathbf{x}'.\mathbf{x}_0)^2 - \frac{1}{2}\mathbf{x}'^2] + O(\frac{1}{r^4})$$

If this result is applied to the potential definition one gets

$$\Phi(\mathbf{x}) = \frac{1}{r}\int\rho d^3\mathbf{x}' + \frac{1}{r^2}\mathbf{x}_0\int\rho\mathbf{x}' d^3\mathbf{x}' + \frac{x_{0_i}x_{0_j}}{2r^3}\int\rho[3x'_ix'_j - x'_n x'_n \delta_{ij}]d^3\mathbf{x}' + O(\frac{1}{r^4})$$

This can be written as well

$$\Phi(\mathbf{x}) = \frac{q}{r} + \frac{\mathbf{p}.\mathbf{x}_0}{r^2} + \frac{Q_{ij}.x_{0_i}x_{0_j}}{2r^3} + O(\frac{1}{r^4})$$

using the definitions

$$q := \int\rho(\mathbf{x}')d^3\mathbf{x}' \qquad \mathbf{p} := \int\mathbf{x}'\rho(\mathbf{x}')d^3\mathbf{x}' \qquad Q_{ij} := \int(3x'_i x'_j - x'_n x'_n \delta_{ij})\rho(\mathbf{x}')d^3\mathbf{x}'$$

**Appendix 2:** decomposition of a general vector field into a potential field and a vortex field

**Theorem 1:**

The derivative of a vector field **F** can be decomposed in a symmetric (index=C) and a antisymmetric part (index=V), i.e.

$$\frac{\partial F_i}{\partial x_j} = \frac{\partial F_i^V}{\partial x_j} + \frac{\partial F_i^C}{\partial x_j}$$

**F$_C$** is the symmetric part and is a gradient of a potential field

$$\frac{\partial F_i^C}{\partial x_j} = \frac{\partial F_j^C}{\partial x_i} \quad or \quad \text{rot } \mathbf{F_C}=0 \quad with \quad F_l^C = \frac{\partial U(x_l)}{\partial x_l}$$

(with $U(x_l)$:=potential function)

**F$_V$** is a antisymmetric vortex field

$$\frac{\partial F_i^V}{\partial x_j} = -\frac{\partial F_j^V}{\partial x_i}$$

Proof:

The derivatives of the field **F** can be decomposed according to

$$\frac{\partial F_i}{\partial x_j} = \frac{1}{2}\left(\frac{\partial F_i}{\partial x_j} + \frac{\partial F_j}{\partial x_i}\right) + \frac{1}{2}\left(\frac{\partial F_i}{\partial x_j} - \frac{\partial F_j}{\partial x_i}\right)$$

for the symmetric part holds:

$$\frac{1}{2}\left(\frac{\partial F_i}{\partial x_j}+\frac{\partial F_j}{\partial x_i}\right) = \frac{1}{2}\left(\frac{\partial F_i^C}{\partial x_j}+\frac{\partial F_i^V}{\partial x_j} + \frac{\partial F_j^C}{\partial x_i}+\frac{\partial F_j^V}{\partial x_i}\right) = \frac{\partial F_i^C}{\partial x_j} = \frac{\partial^2 U}{\partial x_i \partial x_j}$$

for the antisymmetric part holds:

$$\frac{1}{2}\left(\frac{\partial F_i}{\partial x_j}-\frac{\partial F_j}{\partial x_i}\right) = \frac{1}{2}\text{rot}\mathbf{F} = \frac{1}{2}\left(\frac{\partial F_i^C}{\partial x_j}+\frac{\partial F_i^V}{\partial x_j} - \frac{\partial F_j^C}{\partial x_i}-\frac{\partial F_j^V}{\partial x_i}\right) = \frac{\partial F_i^V}{\partial x_j}$$

It can be checked, that

$$\frac{\partial F_i}{\partial x_j} = \frac{\partial F_i^V}{\partial x_j} + \frac{\partial F_i^C}{\partial x_j}$$

q.e.d.

**Theorem 2:**

**F** is a field with a defined boundary condition $\partial F$ around the space which is interesting for the problem. Divergence and rotation are defined according to

$$\nabla \times \mathbf{F} = \mathbf{j}(\mathbf{x}) \qquad \nabla \cdot \mathbf{F} = \rho(\mathbf{x})$$

and the boundary condition $\partial F$

$$\partial F: \quad \mathbf{F} \cdot \mathbf{n} = f(\mathbf{r})$$

Then it holds:

**F** can be calculated as sum of a gradient **F_C** of a potential, plus rotation of a vector potential **F_V**, plus a Laplace field **F_L** according to

$$\mathbf{F} := \mathbf{F_C} + \mathbf{F_V} + \mathbf{F_L}$$

$$\mathbf{F_C} = \frac{1}{4\pi}\int\frac{\rho(\mathbf{x}')(\mathbf{x}-\mathbf{x}')}{|\mathbf{x}-\mathbf{x}'|^3}d^3\mathbf{x}' = -\frac{1}{4\pi}\nabla\int\frac{\rho(\mathbf{x}')}{|\mathbf{x}-\mathbf{x}'|}d^3\mathbf{x}' = -\frac{1}{4\pi}\nabla\Phi$$

$$\mathbf{F_V} = \frac{1}{4\pi}\int\frac{\mathbf{j}(\mathbf{x}')\times(\mathbf{x}-\mathbf{x}')}{|\mathbf{x}-\mathbf{x}'|^3}d^3\mathbf{x}' = \frac{1}{4\pi}\nabla\times\int\frac{\mathbf{j}(\mathbf{x}')}{|\mathbf{x}-\mathbf{x}'|}d^3\mathbf{x}' = \frac{1}{4\pi}\nabla\times\mathbf{A}$$

$$\mathbf{F_L} = \nabla\varphi$$

It holds:

$$\nabla\times\mathbf{F_C} = 0 \qquad \nabla\cdot\mathbf{F_C} = \rho(\mathbf{x})$$
$$\nabla\cdot\mathbf{F_V} = 0 \qquad \nabla\times\mathbf{F_V} = \mathbf{j}(\mathbf{x})$$
$$\nabla\cdot\mathbf{F_L} = \Delta\varphi = 0$$

scheme of the proof [26]:

1) We are interested in the solution of

$$\nabla\times\mathbf{F_C} = 0 \qquad \nabla\cdot\mathbf{F_C} = \rho(\mathbf{x})$$

This is the potential field

$$\mathbf{F_C} = \frac{1}{4\pi}\int\frac{\rho(\mathbf{x}')(\mathbf{x}-\mathbf{x}')}{|\mathbf{x}-\mathbf{x}'|^3}d^3\mathbf{x}' = -\frac{1}{4\pi}\nabla\int\frac{\rho(\mathbf{x}')}{|\mathbf{x}-\mathbf{x}'|}d^3\mathbf{x}'$$

2) We are interested in the solution of

$$\nabla\cdot\mathbf{F_V} = 0 \qquad \nabla\times\mathbf{F_V} = \mathbf{j}(\mathbf{x})$$

This is the vortex field

$$\mathbf{F}_V = \frac{1}{4\pi}\int \frac{\mathbf{j}(\mathbf{x}')\times(\mathbf{x}-\mathbf{x}')}{|\mathbf{x}-\mathbf{x}'|^3} d^3\mathbf{x}' = \frac{1}{4\pi}\nabla\times\int \frac{\mathbf{j}(\mathbf{x}')}{|\mathbf{x}-\mathbf{x}'|} d^3\mathbf{x}'$$

3) We are interested in the solution of

$$\nabla\cdot\mathbf{F}_L = 0 \qquad \nabla\times\mathbf{F}_L = 0$$

using the boundary condition

$$\mathbf{F}_L\cdot\mathbf{n} = \mathbf{F}\cdot\mathbf{n} - \mathbf{F}_C\cdot\mathbf{n} - \mathbf{F}_V\cdot\mathbf{n}$$

The solution is the Laplace field

$$\nabla\cdot\mathbf{F}_L = \Delta\varphi = 0$$

4) The general solution for **F** is the sum of 1) - 3). This can be checked using the vector relations divrot **A**=0 and rotgrad Φ=0 . So one obtains

$$\mathbf{F} = \mathbf{F}_C + \mathbf{F}_V + \mathbf{F}_L$$

q.e.d

The Laplace field is a "generalized constant of integration". It allows to adapt to the boundary conditions. It is needed, if boundary conditions for **F** exist which are non-zero in the infinite, see fig.3.

**Appendix 3:** Derivation of the Lorenz gauge

The continuity equation is

$$\text{div}\,\mathbf{j}(\mathbf{x}') + \frac{d\varrho_E}{dt}(\mathbf{x}') = 0$$

It can be written as

$$\int \frac{\text{div}\,\mathbf{j}(\mathbf{x}') + \dot{\varrho}_E(\mathbf{x}')}{|\mathbf{x}-\mathbf{x}'|} d\mathbf{x}'^3 = 0$$

The divergence term is changed using partial integration. One term can be canceled during partial integration, because $\mathbf{j}(\mathbf{x}')=0$ holds for $\mathbf{x}'= \infty$. So it is obtained

$$\int -\mathbf{j}(\mathbf{x}')\nabla'\frac{1}{|\mathbf{x}-\mathbf{x}'|} + \frac{\dot{\varrho}_E(\mathbf{x}')}{|\mathbf{x}-\mathbf{x}'|} d\mathbf{x}'^3 = 0$$

With $\nabla'|\mathbf{x}-\mathbf{x}'|^{-1} = -\nabla |\mathbf{x}-\mathbf{x}'|^{-1}$ one yields

$$\int \nabla\frac{\mathbf{j}(\mathbf{x}')}{|\mathbf{x}-\mathbf{x}'|} + \frac{\dot{\varrho}_E(\mathbf{x}')}{|\mathbf{x}-\mathbf{x}'|} d\mathbf{x}'^3 = 0$$

This is the Lorenz gauge

$$\nabla\cdot\mathbf{A}_H + \frac{1}{c}\frac{\partial \Phi_E}{\partial t} = 0$$

**Appendix 4:** Model calculations with magnetic monopoles

Model calculation 1:

In order to estimate a field, which could be generated by a magnetic current, we calculate here the non-real case of a coil of one turn which is driven by a current of magnetic charge. The coil is modelled as a rotating tube which is charged with magnetic surface charges.

We use the formulas of magnetostatics applied for magnetic currents (in SI-units) by exchanging the magnetic variables by the analogous electric variables.

Data of the setup:
1 magnetic tube charged with magnetic charges
diameter:                                     d= 2m
height:                                       h= 10cm
number of turns:                              n= 1
magnetic field strength at the surface: $\mathbf{B_0}$ = 1T = 1 Vs/m$^2$
magnetic permeability:                        µ = 10001
speed of rotation:                            f = 10Hz.

Using this data the magnetic current $I_H$ can be calculated to

$$I_H = \text{surface charge} * \text{speed of rotation} = (\mu-1)*\mathbf{B_0}*d*\pi*h*f$$

Then, the electrical field of a magnetic current, cf. **(80)**

$$\mathbf{E} = I_H*n/h = (\mu-1)*\mathbf{B_0}*d*\pi*f = 2*\pi*10^5 \text{ V/m}$$

This means: electrical fields generated by magnetic currents should be sufficiently strong to be detected easily. It should be possible to reach the breakdown voltage of air (30 kV/cm at 1 bar) if the parameters are chosen accordingly high.

Model calculation 2:

We estimate here the field of a permanent magnetic cylinder ring which turns around its central axis. The upper surface of the ring is the north pole, the lower the south pole.

Data of the setup:
1 ring magnet
upper rim: north-, lower rim: south pole

| | |
|---|---|
| diameter | d= 2m |
| height | h= 12cm |
| width | b= 5cm |
| number of turns | n= 1 |
| magnetic field strength at the pole surfaces: | $B_0$= 1T |
| magnetic permeability | µ= 10000 |
| speed of rotation: | f= 10Hz. |

The origin of the coordinate system is the centre of symmetry on the middle of the central axis. The distribution of the electric field lines of the setup can be calculated by using a known example and adapting it for the present setup. For a simple ring current, see fig. 12, Jackson[8] calculates a vortex vector-field in chapter 5.5, equation 5.37. These formula transferred to magnetic currents yields

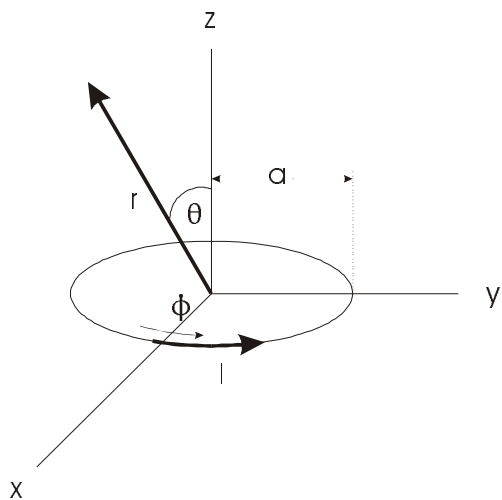

**Fig.12: the coordinate system**

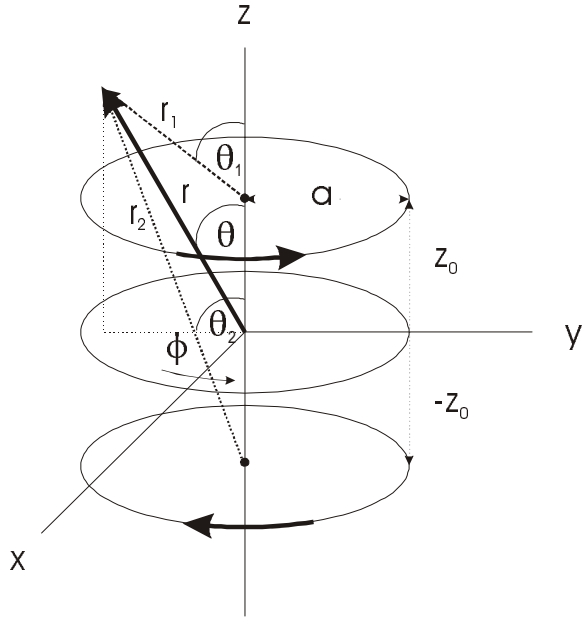

Due to geometry it holds:

z-coordinates: $r_1 \cos\theta_1 + z_0 = r\cos\theta$
$r_2 \cos\theta_2 - z_0 = r\cos\theta$

radius projection into x-y-plane: $r_1 \sin\theta_1 = r\sin\theta$
$r_2 \sin\theta_2 = r\sin\theta$

**fig.13 the geometric situation of a field point due to a circulating magnetic dipol**
the field is composed from two opposite circulating magnetic currents

$$\Gamma_\Phi(r,\theta) = \frac{4 I_H a}{\sqrt{m}} \left[ \frac{(2-m) K(m) - 2 E(m)}{m} \right] \quad \text{with} \quad m = \frac{4 a r \sin(\theta)}{a^2 + r^2 + 2 a r \sin(\theta)}$$

Here are $K(m)$ and $E(m)$ elliptic integrals of first and second order, which are calculated numerically by a program.
This formula is applied for two circuits which are shifted by $z_0$ upward and downwards. In both circuits the magnetic current flows in opposite directions. From the geometry of the setup the appropriate radii and angles of each circuit can be determined, see fig. 13 . The system of equations from fig. 13 are solved

$$\theta_1 = \text{arcot}\left( \frac{r\cos\theta - z_0}{r\sin\theta} \right) \quad \rightarrow \quad r_1 = r\sin\theta / \sin\theta_1$$

$$\theta_2 = \text{arcot}\left( \frac{r\cos\theta + z_0}{r\sin\theta} \right) \quad \rightarrow \quad r_2 = r\sin\theta / \sin\theta_2$$

Then it is possible to write down the electric vortex potential $\Gamma$ which has only one component in $\Phi$-direction which is perpendicular to the plane of the paper, see fig.11

$$\Gamma_\Phi(r,\theta) = \Gamma_\Phi(r_1,\theta_1) + \Gamma_\Phi(r_2,\theta_2)$$

For the purpose of a simple calculation the equally distributed magnetic charge was approximated by 11 charges distributed each over 11 concentric equidistant circuits on the upper and lower surface of the ring. The calculated intensity of the $\Phi$-component of the electric vortex field $\Gamma$ is already shown in fig.10. Then, the **E**-field is calculated according to **E** = rot $\Gamma$, fig. 11 a)+b). So it has been estimated that the field has a maximum of 80 kV/cm near the edges at the surface of the moving magnet. The order of magnitude seems to coincide with the observations of Godin & Roschin [72, 73]. They observed a luminescence and therefore a currents in the ionized air near the surface of moving magnets. According to the usual representation of electrodynamics these currents have to brake down the rotation because any current generated by the induced **E**-field is directed against the original **B**-field due to the Lenz-rule. The authors above, however, report a destabilizing self-acceleration and a weight change at higher angular velocities perhaps due the Brown-Biefeld effect [68]. We mention here that such a self-acceleration could be described by a consistent modification of the theory where all $\rho_H$ are exchanged by $-\rho_H$ in all potential and field expressions. Another possibility are magnetic cycles of the nonlinear magnetic materials.

**Bibliography:**

1. Whittaker, E., *A History of the Theories of Aether and Electricity - The Classical Theories.* Vol. 1. 1973, New York: Humanities Press.
2. Whittaker, E., *A History of the Theories of Aether and Electricity - The Modern Theories 1900 -1926.* Vol. 2. 1973, New York: Humanities Press.
3. Ehrenhaft, F., *Über die Photophorese, die wahre magnetische Ladung und die schraubenförmige Bewegung der Materie in Feldern Erster Teil.* Act. phys. austr., 1951. **4**: p. 461.
4. Ehrenhaft, F., *Über die Photophorese, die wahre magnetische Ladung und die schraubenförmige Bewegung der Materie in Feldern Zweiter Teil.* Act. phys. austr., 1952. **5**: p. 12.
5. Bauer, W.D., *Magnetic Monopoles in Theory and Experiment.* 2002. **2002**.
   URL: http://www.overunity-theory.de/Ehrenhaft/talk.zip
6. Bauer, W.D., *Über elektrische und magnetische Aufladungen - Das Werk von Felix Ehrenhaft*, . 2003: Berlin.
   URL: http://www.overunity-theory.de/Ehrenhaft/Ehrenhaft.zip
7. Priestley, J., *The History and Present State of Electricity, with Original Experiments.* 1767, London.
8. Jackson, J.D., *Classical Electrodynamics Second Edition.* 2 ed. 1975, New York: John Wiley & Sons.
9. Risken, H., *Vorlesungsskript Elektrodynamik.* 1976, Ulm: Uni Ulm Selbstverlag.
10. Oersted, H.C., *Versuche über die Wirkung eines elektrischen Stromes auf die Magnetnadel.* Annals of Philosophy, 1820. **16**: p. 273.
11. Tricker, R.A.R., *Frühe Elektrodynamik.* 1974, Braunschweig: Vieweg.
12. Biot, J.B. and F.Savart, *Eine Bemerkung zum Magnetismus der Volta'schen Säule.* Ann.Chim.Phys., 1820. **15**: p. 222-223.
13. Biot, J.B. and F. Savart, Journal de Physique, 1820. **91**: p. 151.
14. Biot, J.B. and F. Savart, *Magnetisierung von Metallen mittels sich bewegender Elektrizität*, in *Précis Elémentaire de Physique.* 1824: Paris. p. 707-723.
15. Grassmann, H., *Neue Theorie der Elektrodynamik.* Poggendorf's Annalen der Physik und Chemie, 1845. **64**(1): p. 1 - 18.
16. Ampère, A.M., Ann. Chim. Phys., 1820. **15**: p. 59-76, 177-208.
17. Ampère, A.M., Mém. de l'Acad., 1825. **6**: p. 175.
18. Riemann, B., *Schwere, Elektrizität und Magnetismus, nach den Vorlesungen von B.Riemann.* 1875, Hannover.
19. Cavallieri, G., G. Spavieri, and G. Spinelli, *The Ampère and Biot-Savart force laws.* Eur. J. Phys., 1996. **17**: p. p.205-207.
20. Curé, J.C., *Action and reaction in Electrodynamics.* Deutsche Physik, 1995. **4**(13): p. 5 - 10.
21. Aspden, H., *Physics without Einstein.* 1969, Southampton: Sabberton.
22. Marinov, S., *Marinov's formula is the only viable formula in magnetism.* Deutsche Physik, 1994. **3**(11): p. 18 - 34.
23. Cavallieri, G., *et al.*, *Experimental proof of standard electrodynamics by measuring the self-force on a part of a current loop.* Phys. Rev. E, 1998. **58**(2):

p. 2502 - 2517.
24. Neumann, F.v., *Abhandlungen I*. 1845, Berlin.
25. Neumann, F.v., *Abhandlungen II*. 1848, Berlin.
26. Bronstein, I.N. and K.A. Semendjajew, *Taschenbuch der Mathematik*. 20 ed. 1983, Thun: Harri Deutsch.
27. Thomson, J.J., Phil. Mag., 1899. **48**(5): p. 547.
28. Graneau, P., *Comment on "The motionally induced back EMF in railguns"*. Phys. Lett. A, 1991. **160**: p. 490 - 491.
29. Graneau, P. and N. Graneau, *Newtonian Electrodynamics*. 1995, New York: World Scientific Publishing Co.
30. Tanberg, R., *On the cathode of an arc drawn in vacuum.* Phys. Rev, 1929. **35**: p. 1080.
31. Kobel, E., *Pressure and high velocity vapour jets at cathodes of a mercury vacuum arc.* Phys. Rev., 1930. **36**: p. 1636.
32. Johansson, L., *Longitudinal electrodynamic forces - and their possible technological application*, in *Department of Electromagnetic Theory*. 1996, Lund Institute of Technology P.O. Box 118 S-22100 Schweden: Lund.
   URL: http://www.df.lth.se/~snorkelf/LongitudinalMSc.pdf
33. Rambaut, M. and J.P. Vigier, *Ampère forces considered as collective non-relativistic limit of the sum of all Lorentz interactions acting on individual current elements: possible consequences for electromagnetic discharge stability and tokamak behaviour.*
   Phys. Lett. A, 1990. **148**(5): p. 229-238.
34. Beauregard, O.C.d., *Statics of filaments and magnetostatics of currents: Ampère tension and the vector potential.* Phys. Lett. A, 1993. **183**: p. 41 - 42.
35. Duschek, A. and A. Hochrainer, *Grundzüge der Tensorrechnung in analytischer Darstellung 2. Tensoranalysis*. Vol. 2. 1950, Berlin: Springer-Verlag.
36. Duschek, A. and A. Hochrainer, *Grundzüge der Tensorrechnung in analytischer Darstellung 1. Tensoralgebra*. 3. Auflage ed. Vol. 1. 1954, Berlin: Springer-Verlag.
37. Mikhailov, V.F. and L.I. Mikhailova, *Preprint HEPI-82-01*, . 1982, Academy of Sciences of Kazakh SSR: Alma Ata.
38. Mikhailov, V.F., *The Magnetic Charge Phenomen on Ferromagnetic Aerosols.* Phys. Lett. B, 1983. **130**: p. 331.
39. Mikhailov, V.F., J. Phys. A: Math. Gen., 1985. **18**: p. L903.
40. Mikhailov, V.F., Ann. Fond. Louis de Broglie, 1987. **12**: p. 491.
41. Mikhailov, V.F. and L.I. Mikhailova, *On some regularities of aerosol particle motion in electromagnetic fields*, . 1987, Kazakh Akademy of Sciences, USSR: Alma Ata.
42. Mikhailov, V.F., *Preprint HEPI-88-17*, . 1988, Academy of Sciences of Kazakh SSR: Alma Ata.
43. Mikhailov, V.F., *Preprint HEPI-88-05*, . 1988, Academy of Sciences of Kazakh SSR: Alma Ata.
44. Mikhailov, V.F. and L.I. Mikhailova, J. Phys. A: Math. Gen., 1989. **23**: p. 53 - 63.
45. Mikhailov, V.F. and L.I.Mikhailova, *Preprint HEPI-90-07*, . 1990, Academy of Sciences of Kazakh SSR: Alma Ata.
46. Mikhailov, V.F. and L.I.Mikhailova, *Preprint HEPI-91-04*, . 1991, Academy of

Sciences of Kazakh SSR: Alma Ata.
47. Mikhailov, V.F., *Observation of apparent magnetic charges carried by ferromagnetic particles in water droplets.* J. Phys. A: Math. Gen., 1991. **24**: p. 53 - 57.
48. Mikhailov, V.F., *Experimental detection of Dirac's magnetic charge ?* J. Phys. D: Appl. Phys., 1996. **29**: p. 801 - 804.
49. Tricker, R.A.R., *Faraday und Maxwell.* 1974, Berlin: Akademie-Verlag.
50. Obukhov, Y.N. and F.W. Hehl, *Electromagnetic energy-momentum and forces in matter.* Phys. Lett. A, 2003. **311**: p. 277 - 284.
51. Rainich, G.Y., *Electrodynamics in the General Relativity Theory.* Trans. Am. Math. Soc., 1925. **27**(1): p. 106 - 136.
52. Katz, E., *Concerning the Number of Independent Variables of the Classical Electromagnetic Field.* Am. J. Phys., 1965. **33**: p. 306 - 312.
53. Harrison, H., et al., *Possibility of Observing the Magnetic Charge of an Electron.* Am. J. Phys., 1963. **31**: p. 249.
54. Akers, D., *Mikhailov´s Experiments on Detection of Magnetic Charge.* Int. J. Theor. Phys., 1988. **27**(8): p. 1019 - 1022.
55. Dirac, P.A.M., *Quantizised Singularities in the Electromagnetic Field.* Proc. R. Soc. London A, 1931. **133**: p. 60.
56. Dirac, P.A., *The Theory of Magnetic Poles.* Phys. Rev., 1948. **74**(7): p. 830.
57. Landau, L.D. and E.M. Lifshitz, *Lehrbuch der theoretischen Physik - Elektrodynamik der Kontinua.* 5 ed. Vol. 8. 1990, Berlin: Akademie-Verlag.
58. O´Dell, T.H., *The Electrodynamics of Magneto-electric Media.* Series of monographs on selected topics in solid state physics. 1970, London: North-Holland Amsterdam.
59. Penfield, P. and H. Haus, *Electrodynamics of moving media.* 1967, Cambridge Mass.
60. Hubert, A. and R.Schäfer, *Magnetic Domains.* 2001, Berlin,Heidelberg: Springer-Verlag.
61. Ternovsky, V., B. Luk'yanchuk, and J.P. Wang, *Remanent States of Small Ferromagnetic Cylinder.* JETP Letters, 2001. **73**(12): p. 661 - 665.
62. Bauer, W.D., *Second Law versus Variation Principles*, . 2001.
URL: http://xxx.lanl.gov/pdf/physics/0009016.pdf
63. Christoph, J., *Musterbildung auf Elektrodenoberflächen*, in *Chemie*, FU Berlin: Fachbereich Chemie, Berlin., 2000
64. Yusa, G. and H. Sakaki, *Trapping of photogenerated carriers by InAs quantum dots and persistent photoconductivity in novel GaAs/n-AlGaAs field-effect transistor structures.* Appl. Phys. Lett., 1997. **70**(3): p. 345 - 347.
65. Rack, A., et al., *Dynamic bistability of quantum dot structures: Role of the Auger process.* Phys.Rev. B, 2002. **66**: p. 165429.
see as well URL: http://www.hmi.de/people/rack/diplom/index.html
66. Rego, L.G.C. and G. Kirczenow, *Electrostatic mechanism for semiconductor heterostructures.* Appl. Phys. Lett., 1999. **75**(15): p. 2262 -2264.
67. Ehlers, H.J., *raum&zeit-Interview mit John Roy Robert Searl*, in *Raum&Zeit special*, H.J. Ehlers, Editor. 1994, Ehlers-Verlag: Sauerlach. p. 148 - 149.
68. Sandberg, S.G., *Der Searl-Effekt und der Searl-Generator*,
in *Raum&Zeit special 7 Wunschtraum der Menschheit: Freie Energie*,

H.J. Ehlers, Editor. 1994, Ehlers Verlag: Sauerlach. p. 149 - 157.
see as well URL: http://www.rexresearch.com

69. Schneider, H. and H. Watt, *Dem Searl-Effekt auf der Spur (I)*, in *Raum&Zeit special*, H.J. Ehlers, Editor. 1994, Ehlers Verlag: Sauerlach. p. 174 - 180.
70. Schneider, H. and H. Watt, *Dem Searl-Effekt auf der Spur (II)*, in *Raum&Zeit special*, H.J. Ehlers, Editor. 1994, Ehlers-Verlag: Sauerlach. p. 181 - 185.
71. Manning, J. and P. Sinclair, *The Granite Man and the Butterfly*. 1995, Fort Langley, B.C. , Canada: Project Magnet Inc.
see as well URL: http://jnaudin.free.fr/html/hammnu.htm
72. Roschin, V.V. and S.M. Godin, *An Experimental Investigation of the Physical Effects in a Dynamic Magnetic System.* Technical Physics Letters, 2000. **26**(12): p. 1105 - 1107.
see as well URL: http://www.rexresearch.com/roschin/roschin.htm
or Russian patent no. RU2,155,435 Publication date: 2000-08-27
73. Roschin, V.V. and S.M. Godin, *Experimental Research of the Magnetic-Gravity Effects*, . 2000.
see as well URL: http://alexfrolov.narod.ru/russearl.html
74. Bruno, P., *Nonquantized Dirac Monopoles and Strings in the Berry Phase of Anisotropic Spin Systems*, Phys.Rev.Lett. 2004. 93: 247202
75. N.N. , *Der Einstein-deHaas Effekt*, Praktikumsanleitung TU-Karlsruhe 5.12.2001
http://www.uni-karlsruhe.de/3Block1.php/Studium/F-Praktika/Downloads/de-Haas.pdf